\documentclass[twocolumn]{aastex63_rt42}

\usepackage{graphicx}
\usepackage{epstopdf}
\usepackage{amsmath}
\usepackage{multirow}
\usepackage{color}
\usepackage{icomma}
\usepackage{groundtodust}

\def\um{\mu\textrm{m}}

\def\cm{\textrm{cm}}
\def\meter{\textrm{m}}

\def\erg{\textrm{erg}}
\def\ev{\textrm{eV}}

\def\Watt{\textrm{W}}

\def\Kelvin{\textrm{K}}

\def\yr{\textrm{yr}}
\def\kyr{\textrm{kyr}}
\def\Myr{\textrm{Myr}}
\def\Gyr{\textrm{Gyr}}

\def\sigmaSB{\sigma_{\rm SB}}

\def\Msun{{M}_{\odot}}
\def\Lsun{{L}_{\odot}}
\def\Rsun{{R}_{\odot}}
\def\rhosun{\rho_{\odot}}
\def\MEarth{{M}_{\oplus}}

\def\REarth{{R}_{\oplus}}

\def\gram{\textrm{g}}

\def\rhoUnits{\textrm{g}\,\textrm{cm}^{-3}}

\def\eV{\textrm{eV}}

\def\AU{\textrm{AU}}
\def\pc{\textrm{pc}}

\def\SigmaUnits{\textrm{g}\,\textrm{cm}^{-2}}
\def\kms{\textrm{km\,s}^{-1}}

\def\ga{\gtrsim}
\def\la{\lesssim}

\def\endash{\text{--}}

\newcommand{\editOne}[1]{#1}

\shorttitle{Megastructures and Collisional Cascades}
\shortauthors{Lacki}

\begin{document}

\title{Ground to Dust: Collisional Cascades and the Fate of Kardashev II Megaswarms}

\correspondingauthor{Brian C. Lacki}
\email{astrobrianlacki@gmail.com}
\author[0000-0003-1515-4857]{Brian C. Lacki}
\affiliation{Breakthrough Listen, Department of Physics, Denys Wilkinson Building, Keble Road, Oxford OX1 3RH, UK }

\begin{abstract}
Extraterrestrial intelligences are speculated to surround stars with structures to collect their energy or to signal distant observers. If they exist, these most likely are megaswarms, vast constellations of satellites (elements) in orbit around the hosts. Although long-lived megaswarms are extremely powerful technosignatures, they are \editOne{liable} to be subject to collisional cascades once guidance systems start failing. The collisional time is roughly an orbital period divided by the covering fraction of the swarm. Structuring the swarm orbits does not prolong the initial collisional time as long as there is enough randomness to ensure collisions, although it can reduce collision velocities. I further show that once the collisional cascade begins, it can develop extremely rapidly for hypervelocity collisions. Companion stars or planets in the stellar system induce perturbations through the Lidov-Kozai effect among others, which can result in orbits crossing within some millions of years. Radiative perturbations, including the Yarkovsky effect, also can destabilize swarms. Most megaswarms are thus likely to be short-lived on cosmic timescales without active upkeep. I discuss possible mitigation strategies and implications for megastructure searches.
\end{abstract}

\keywords{Search for extraterrestrial intelligence --- Technosignatures --- Collisional processes --- Astrodynamics --- Orbits}

\section{Introduction}
\label{sec:Intro}
The Search for Extraterrestrial Intelligences (SETI) must deal with the challenge that technological societies are separated not just in space, but in time. Even if every single star is home to intelligent life at some point during its lifespan, the odds are against us observing them unless the lifespan of the technosignatures is many millions of years. But if they are short-lived, might some technosignatures greatly outlast their creators? The long-term survival of actively broadcasting transmitters is problematic without maintenance -- especially ones visible over interstellar distances. Radio transmitters, lasers, and broadcasting probes all presumably require complicated electronics, or even more advanced technology, any of which could eventually break down, and the former two need to avoid \editOne{failure} while channeling large amounts of power \citep[e.g.,][]{Sheikh20}. Of course, we can speculate that such technology could be self-repairing, or even self-replicating, but arguably the builders' society is itself a self-repairing mechanism; if even that cannot survive for so long, why should their broadcasting technology?

Technosignatures that are rarely produced but long-lived could outnumber those that are universal but short-lived, and thus have a greater chance of being observable \citep{Kipping20,Balbi21,Lacki24}. A passive technosignature, one that remains detectable without any maintenance simply by virtue of continuing to exist, would bridge the gap, serving as a fossil of its creators \citep{Carrigan12}. A probe within our Solar System could be a robust technosignature even if it is derelict (\citealt{Rose04}; see also \citealt{Lacki19-Glint}). Because they are not operating under their own power, most conceivable passive technosignatures would be beyond detection at interstellar distances. But the megastructures, a hypothetical class of artificial objects as big as planets, stars, or even larger, could make themselves known over vast distances. It is tempting to imagine such structures as nigh indestructible, so large that any disaster would be a mere scratch, and that once built, they exist forever. If so, then megastructures bridge both the enormous gulfs of space and time that SETI must overcome.

The archetypal megastructure is the Dyson sphere, a shell that completely encloses a star, gathering all its power and providing exponentially more surface area than planets \citep{Dyson60,Wright20}. The concept has been remixed many times over the decades -- structures with different functions like maximizing computation \citep{Bradbury00,Wright23} or altering the host's evolution \citep{Criswell85,Huston22,Scoggins23}; partial structures that only cover a fraction of the available sightlines or involve translucent swarms \citep{Zackrisson18,Suazo22}; placing them around different objects like black holes \citep{Inoue11,Semiz15,Osmanov16,Hsiao21}; and scaling down or up its size. On the small end are thin shapes the size of planets but of varying shapes, which could be used to signal the presence of an ETI as their odd shapes distort the transit light curve \citep{Arnold05,Wright16}. Modulation of pulsar light curves is another possibility \citep{Chennamangalam15}. Perhaps the largest solid structure proposed is described in \citet{Kardashev85}: a single parsecs-wide disk built around a galactic nucleus. Nor are megastructures necessarily isolated oddities: the resources to embark on interstellar voyages are negligible compared to that to remake a stellar system; perhaps there are entire galaxies out there whose stellar populations are cloaked by these objects (\citealt{Wright14-Paradox}; but see \citealt{Lacki19-Sunscreen}). 

Now, a single solid structure the size of a solar system is thought impractical with our current understanding of material sciences. The stresses would be too great for one to ever be built, and they suffer from gravitational instabilities \citep{Wright20}. Most conceptions of megastructures bigger than planets picture them as dense swarms, each element built with conventional materials, but so numerous that the optical depth can exceed one. The swarm elements orbit the host sun like any other body. The orbiting swarm concept is very general; it can be scaled up to galactic scales with the galactic ``blackboxes'' of \citet{Lacki16-K3}, in which the elements are artificial dust grains in an interstellar medium. It can also be scaled down to dense swarms of satellites orbiting a planet \citep{SocasNavarro18,Sallmen19}. An orbiting planetary swarm has the advantage of being feasible and precedented: the growing enormous constellations of satellites in Earth orbit can be viewed as such a swarm in the earliest stages of development. Even in the earliest days of the Space Age, the West Ford Project unilaterally spread millions of copper dipoles in medium Earth orbit to radically alter the planet's radio environment \citep{Morrow61,Shapiro66}.\footnote{Both the present day satellite swarms and the West Ford Project sparked protest from astronomers worried about their impacts on radio and optical astronomy \citep[e.g.,][]{Lovell62,Sandage63,Hainaut20,McDowell20}.} The Kardashev scale, inspired by \citet{Kardashev64}, presents a natural categorization of swarms: Type I for planetary, Type II for stellar, and Type III for galactic.\footnote{The Kardashev scale as defined in \citet{Kardashev64} actually measures the power used in a broadcast; I am using it in the common very loose sense of the scale of a technosignature.}

Herein lies a major threat to megastructure lifespans: if abandoned, the individual elements eventually start crashing into each other at high speeds \citep[as noted in][]{Lacki16-K3,Sallmen19,Lacki20-Lenses}. Not only do the collisions destroy the crashed swarm members, but they spray out many pieces of wreckage. Each of these pieces is itself moving at high speeds, so that even pieces much smaller than the original elements can destroy them. Thus, each collision can produce hundreds of missiles, resulting in a rapid growth of the potentially dangerous population and accelerating the rate of collisions.  The result is a collisional cascade, where the swarm elements are smashed into fragments, that are in turn smashed into smaller pieces, and so on, until the entire structure has been reduced to dust. Collisional cascades are thought to have shaped the evolution of minor Solar System body objects like asteroid families and the irregular satellites of the giant planets \citep{Kessler81,Nesvorny03,Nesvorny04,Bottke10}. They also thin out debris disks, both in and out of the Solar System \citep{Wyatt08,Nesvorny10}. The specter of the technological collisional cascade was brought up by \citet{Kessler78}, which noted that space debris from our activities in low Earth orbit posed a threat to operating satellites. This threat has only grown with the deployment of large satellite constellations and anti-satellite missile tests \citep{LeMay18,Boley21}, and this for a planetary swarm with an optical depth far less than one. In a structure as dense as a Dyson sphere, the effects can be immediate.

\subsection{The megaswarms considered}
Megaswarms are characterized by having a great many potentially large objects in orbits around the host, but otherwise their conceptions are very diverse. Both element area and numerosity directly increase the difficulty of avoiding collisions. I will focus on two cases, which might be taken as the limits of a dense and rarefied megaswarm. 

The prototypical megaswarm, the Dyson sphere, emphasizes the numerosity of swarm elements -- whatever their individual size, there have to be enough to cover the star in all directions, implying a literally astronomical number of elements. This is to collect as much energy as possible from the host sun. Even a partial Dyson sphere will have a great many. Originally, the elements were considered to be inhabited space stations, and thus might be kilometers wide \citep{Dyson60}. A later conception suggests that they might be used simply to power computers, in which case \editOne{their sizes} might be \editOne{arbitrary \citep[e.g.,][]{Bradbury00,Bradbury11,Wright23}}. Because of this uncertainty in the size, it is more useful in this paper to consider the mean optical depth of the swarm, $\Mean{\sOptdepth} \sim \Mean{\esDensity} \Mean{\eArea} \Mean{\sRin}$, where $\esDensity$ is the number density of elements in the swarm, $\eArea$ is the projected area of each element, and $\sRin$ is the thickness of the shell. When $\sOptdepth \la 1$, it is essentially the covering factor. A full Dyson swarm is discussed in this paper as a maximal megaswarm.

On the other hand, an occulter swarm simply places objects detectable by their transit light curves, which can give clues to their artificial natures \citep{Arnold05}. This means that the individual elements are as wide as planets (though presumably much thinner, more like plates or thin bubbles), but there may be relatively few of them. But like a transiting exoplanet, an artificial occulter has a problem as ``beacons'' meant to draw notice to a planetary system: they are seen from only a small range of inclinations (``the transit zone''; \editOne{cf.}, \citealt{Heller16,Wells18}). The transit zone has an angular width of $2 \hR / \sRout$, for an element that transits a star of radius $\hR$ at a distance $\sRout$. Since the range of inclinations is $\pi$ radians, this means that the number of elements in the swarm must be
\begin{equation}
\esN \ga \frac{\pi \sRout}{2 \hR} \approx 340 \left(\frac{\sRout}{\AU}\right) \left(\frac{\hR}{\Rsun}\right)^{-1}
\end{equation}
to be effective as a beacon along all directions. An occulter swarm with this many elements is referred to herein as the \emph{minimal occulter swarm}, and will be considered as the minimal megaswarm. Actual occulter swarms may have more elements, to account for the orbital wander of the elements over time from perturbations (Section~\ref{sec:Perturb}), for especially attention-grabbing transit signatures \citep{Arnold05}, or so that transits are more frequent than once per orbital period. \citet{Lacki20-Lenses} posited that ETIs might use lenses around accreting neutron stars to create bright beams visible as X-ray flashes; this construction too would be denser than a minimal occulter.

\subsection{Outline}
This paper lays out some dangers of collisional cascades for megastructure swarms, with a particular focus on the Kardashev type II stellar swarms. I start by presenting estimates of the initial collisional destruction time and discuss the likely geometry of a megaswarm (Section~\ref{sec:CollisionOverview}). The actual process of collisional cascades are modeled in Section~\ref{sec:Cascades}, including the ultimate fate of the dust. Section~\ref{sec:Perturb} discusses the many possible sources of gravitational perturbations that can lead to orbit crossing, with a focus on the Lidov-Kozai effect. Radiative thrust, both as a source of lift and as a destabilizing force, is discussed in section~\ref{sec:Radiative}. The final discussion section (\ref{sec:Implications}) lays out the circumstances in which a megastructure could survive for a very long time: what sorts of structures around what sorts of objects, and the implications for SETI. Section~\ref{sec:Conclusions} summarizes the conclusions.

\section{Collision time estimates and the structure of megaswarms}
\label{sec:CollisionOverview}
When discussing \editOne{collisional} cascades, we must make a distinction between two different timescales. The first, treated in this section, is called here the \emph{collisional} timescale $\xTColl$. It is the inverse of the rate at which a megaswarm element collides with other objects: if the swarm around the element remains the same, we expect it to have about one collison over $\xTColl$. As used here, the collisional timescale is an instantaneous quantity defined for the initial population of the swarm.

The other timescale is the \emph{cascade} timescale $\xTCasc$, which describes the actual time it takes for collisions to grind the swarm down to fragments and is dealt with in section~\ref{sec:Cascades}. This can be much faster than the collisional timescale, because early collisions multiply the number of potential missiles in the system -- as the cascade develops, the environment of an element does \emph{not} remain the same. Still, the two are linked: the cascade's initial growth is a result of the first collisions in the swarm, when the conditions are the same as those used for the collisional time. The collisional time is kinematic, while the cascade time is evolutionary.

\subsection{The collisional time}

\begin{figure}
\centerline{\includegraphics[width=9cm]{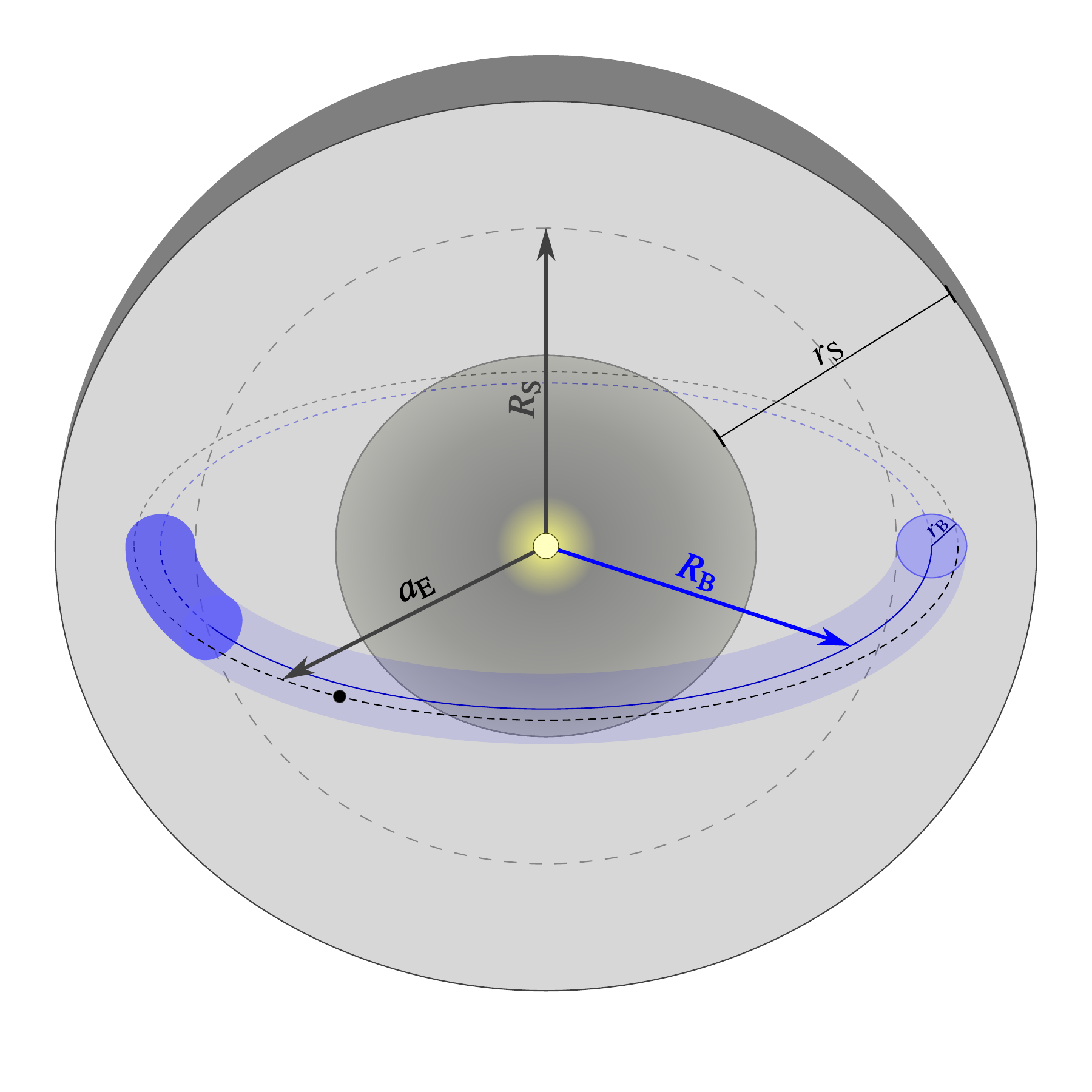}}
\figcaption{Sketch of the assumed shell geometry of a megaswarm. The shell contains belts, one example of which is depicted in blue. Within each belt, elements orbit, possibly slightly displaced from the belt center by random velocity deviations (black dashed line). Note that the belts do not fill the entire swarm shell, much of which may be empty.\label{fig:Geometry}}
\end{figure}

The naive collisional time is based on the assumption that the swarm is completely randomized, with each element experiencing a constant flux of projectiles, each independent of the last. Let us suppose the elements are mixed into a swarm volume, an isotropic shell with mean radius $\sRout$ and thickness $\sRin$, for a volume $\sVolume \approx 4 \pi \sRout^2 \sRin$ (Figure~\ref{fig:Geometry}). The velocity dispersion of the swarm elements is $\sDispV$ and the mean collisional cross section between elements is $\eeArea$.\footnote{This collisional cross section is different than the geometric or projected area $\eArea$. Two spherical elements of equal radius $\eRadius$ collide if one comes within $2\eRadius$ of the other, so the collisional cross section is $4 \pi \eRadius^2$, not $\pi \eRadius^2$. Of course, collisions at the maximum range are glancing, and may have different physics, but that is beyond the scope of this work.}
 For a swarm with $\esN$ elements, the collisional time is
\begin{equation}
\sTCol \approx \frac{\sVolume}{\esN \sDispV \eeArea} \approx \frac{4 \pi \sRout^2 \sRin}{\esN \sDispV \eeArea}.
\end{equation}
The circular orbital velocity, $\xVCirc(\SemimajorCore) \equiv \sqrt{G_N \hM / \SemimajorCore}$ for a body with semimajor axis $\SemimajorCore$, taken at the swarm location, is a characteristic velocity scale. And indeed, if elements of all different inclinations, or widely varying eccentricities, are all mixing together, the velocity dispersion should be of this magnitude. The assumption that $\sDispV \approx \xVCirc(\sRout)$ gives us the naive collisional time,
\begin{equation}
\label{eqn:NaiveCollisional}
\sTColNAIVE \equiv \frac{4 \pi \sRout^2 \sRin}{\esN \eeArea \xVCirc(\sRout)}.
\end{equation}

We can then write the collisional time in terms of the characteristic orbital period, $\ePeriod(\sRout) \approx 2 \pi \sRout / \xVCirc(\sRout)$, and swarm covering fraction, $\sFcover \approx \esN \eArea / (4 \pi \sRout^2)$ \citep[compare][\editOne{equation 16}]{Wyatt99}:
\begin{equation}
\sTCol \approx \ePeriod(\sRout) \cdot \frac{1}{2 \pi \sFcover} \frac{\sRin}{\sRout} \frac{\xVCirc(\sRout)}{\sDispV} \frac{\eArea}{\eeArea}. 
\end{equation}
In the limit of a fully randomized Dyson swarm ($\sFcover \sim 1$, $\sDispV/\xVCirc(\sRout) \sim 1$), then, the collisional time is basically the orbital period; the swarm self-destructs within a year or so. Previous studies have noted the need for course correction in Dyson swarms to inhibit gravitational instabilities \citep{Wright20}, but since the dynamical time is $\sim 1/\sqrt{G_N \sMDensity}$ for a swarm mass density $\sMDensity$, these should take about a thousand orbital periods to develop (assuming a mass $\sim 10^{-6}$ of the host, like a terrestrial planet might \editOne{have}). Collisional cascades pose a \emph{much} more urgent hazard.

\begin{widetext}
The naive collisional time for a minimal occulter swarm is:
\begin{equation}
\sTColNAIVE \approx \frac{8 \sRout^{5/2} \hR}{G_N^{1/2} \hM^{1/2} \eeArea} \left(\frac{\sRin}{\sRout}\right) \approx 1.0\ \Myr \left(\frac{\sRout}{\AU}\right)^{5/2} \left(\frac{\hM}{\Msun}\right)^{-1/2} \left(\frac{\eeArea}{\pi \REarth^2}\right)^{-1} \left(\frac{\hR}{\Rsun}\right) \left(\frac{\sRin}{\sRout}\right) .
\end{equation}
While certainly long-lived, it is still vastly shorter than the lifespan of a star unless the occulter swarm is very far from the star or the occulters are very small; and even if every solar analog hosted one of these swarms at 1 AU once during its lifespan, only one in twenty thousand might be expected to still have one unless they are actively maintained.\footnote{When transiting an inner binary, the stars themselves move and the minimal occulter model no longer applies, but more elements still means more frequent transits \citep{Lacki20-Lenses}.}
\end{widetext}

Table~\ref{table:CollisionTimes} gives some characteristic collisional times for megaswarms experiencing Earth-like \editOne{instellations} around various host stars.

\begin{deluxetable*}{llcccccccccccc}
\tabletypesize{\footnotesize}
\tablecolumns{13}
\tablewidth{0pt}
\tablecaption{Characteristic collisional and cascade times for habitable megaswarms \label{table:CollisionTimes}}
\tablehead{\colhead{Host type} & \multicolumn{3}{c}{Host properties} & \multicolumn{4}{c}{Swarm properties} & \multicolumn{2}{c}{Dyson swarm} & \multicolumn{3}{c}{Minimal occulter} \\
& \colhead{$\hL$} & \colhead{$\hM$} & \colhead{$\hR$} & \colhead{$\sRout$} & \colhead{$\sVCirc$} & \colhead{$\ePeriod$} & \colhead{$\xVPARAM$} & \colhead{$\sTColNAIVE$} & \colhead{$\sTCascEST$} & \colhead{$\esN$} & \colhead{$\sTColNAIVE$} & \colhead{$\sTCascEST$} \\
 & \colhead{($\Lsun$)} & \colhead{($\Msun$)} & \colhead{($\Rsun$)} & \colhead{($\AU$)} & \colhead{($\kms$)} & \colhead{($\yr$)} & & \colhead{($\yr$)} & \colhead{($\yr$)} & & \colhead{($\yr$)} & \colhead{($\yr$)}}
\startdata      
M dwarf          & $0.001$  & $0.2$ & $0.1$  & $0.032$ & $75$  & $0.013$  & $240$ & $0.0020$  & $1.8 \times 10^{-5}$ & $11$     & $41$                 & $0.36$\\
K dwarf          & $0.1$    & $0.7$ & $0.7$  & $0.32$  & $44$  & $0.21$   & $140$ & $0.034$   & $0.00070$            & $110$    & $49000$              & $1000$\\
Solar twin       & $1$      & $1$   & $1$    & $1.0$   & $30$  & $1.0$    & $94$  & $0.16$    & $0.0063$             & $340$    & $1.0 \times 10^6$    & $41000$\\
Vega twin        & $50$     & $2$   & $2.5$  & $7.1$   & $16$  & $13$     & $50$  & $2.1$     & $0.22$               & $2400$   & $2.4 \times 10^8$    & $2.6 \times 10^7$\\
Red giant        & $200$    & $1$   & $25$   & $14$    & $7.9$ & $53$     & $25$  & $8.5$     & $2.3$                & $4800$   & $2.0 \times 10^{10}$ & $5.3 \times 10^9$\\
Red supergiant   & $100000$ & $20$  & $700$  & $320$   & $7.5$ & $1300$   & $24$  & $200$     & $58$                 & $110000$ & $2.9 \times 10^{14}$ & $8.4 \times 10^{13}$\\
White dwarf      & $0.0001$ & $0.6$ & $0.01$ & $0.01$  & $230$ & $0.0013$ & $730$ & $0.00021$ & $2.8 \times 10^{-7}$ & $3.4$    & $0.13$               & $0.00018$
\enddata
\tablecomments{The radius $\sRout$ of the swarm is at the distance where the \editOne{instellation} is the same as the Earth gets from the Sun, with $\sRin = \sRout$. Each swarm is assumed to be fully randomized ($\sVCirc \equiv \xVCirc(\sRout) = \sDispV$). For the Dyson swarm, the mean optical depth is taken to be $1$; for the minimal occulter swarm, each element has a projected area equal to that of the Earth. The elements in each swarm have an impact strength of $10^9\ \erg\,\gram^{-1}$.}
\end{deluxetable*}

\subsection{The packing of orbital belts in swarms}

The basic estimates of the collisional time may seem unrealistically short. Surely the builders would not scatter the element's orbits at random, knowing about the threat. And indeed, such issues have already begun to be considered for humanity's own little swarms, the satellite constellations. If none of the orbits of the swarm elements intersect, then no collisions can occur. This is why the inner Solar System has not self-destructed in geological time, because each of the inner planets has ``cleared its orbit'' during the process of terrestrial planet formation. Only if the orbits change drastically, as might be expected over billions of years, would we have to worry about them running into one another. But while it is easy to fit the orbits of four planets (and the Moon) into the inner Solar System without crossing one another, would that be true in an occulter swarm with hundreds of planet-sized bodies? What about a full-on Dyson swarm with many billions of elements? 

\begin{figure*}
\centerline{\includegraphics[width=18cm]{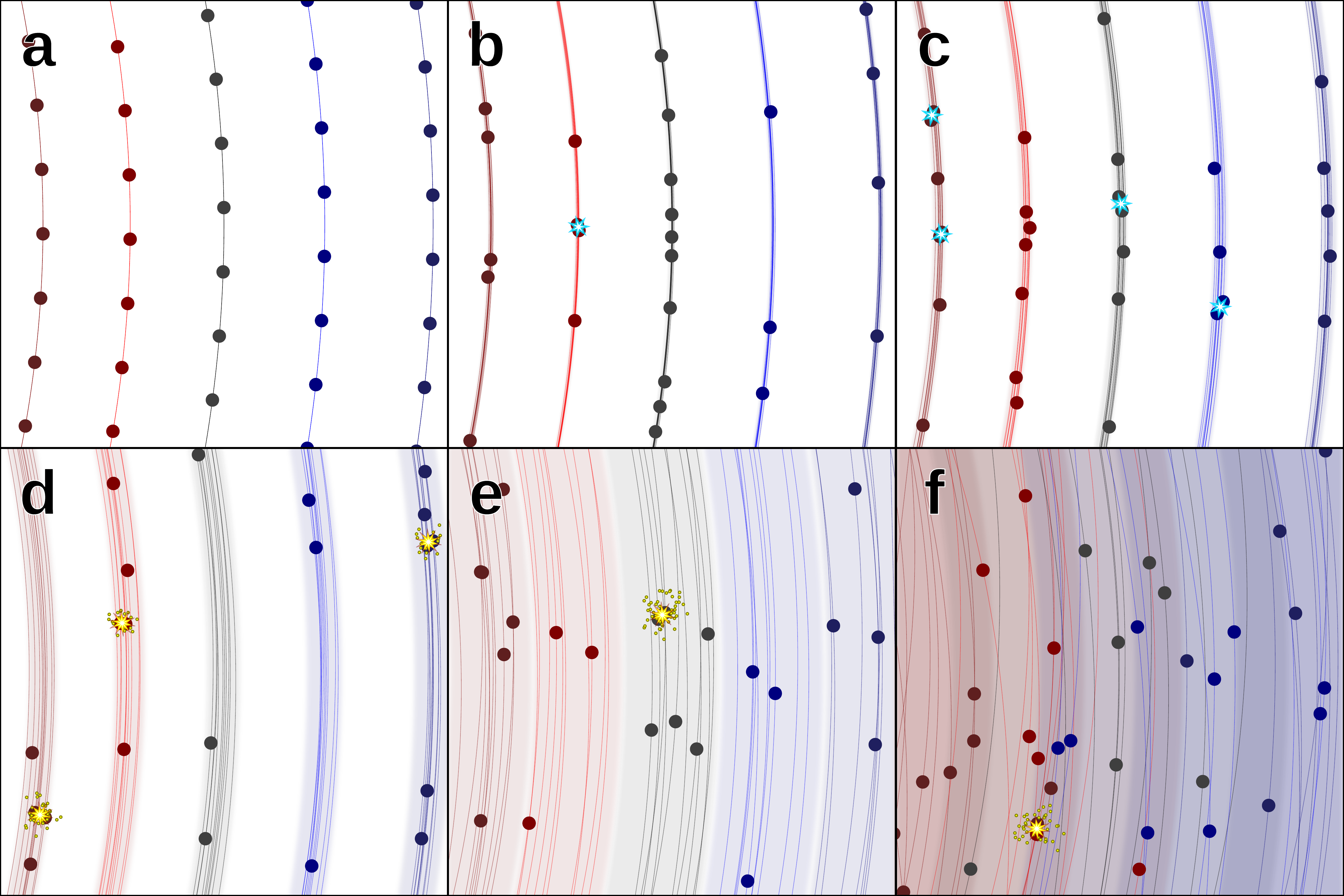}}
\figcaption{A sketch of a series of coplanar belts heating up with randomized velocities. In panel (a), the belt is a single orbit on which elements are placed in an orderly fashion. Very small random velocities (meters per second or less) cause small deviations in the elements' orbits, though so small that the belt is still ``sharp'', narrower than the elements themselves (b). The random velocities cause the phases to desynchronize, leading to collisions, although they are too slow to damage the elements (cyan bursts). The collision time decreases rapidly in this regime until the belt is as wide as the elements themselves and becomes ``fuzzy'' (c). The collision time is at its minimum, although impacts are still too small to cause damage. In panel (d), the belts are still not wide enough to overlap, but relative speeds within the belts have become fast enough to catastrophically damage elements (yellow explosions), and are much more frequent than the naive collisional time implies because of the high density within belts. Further heating causes the density to fall and collisions to become rarer until the belts start to overlap (e). Finally, the belts grow so wide that each belt overlaps several others, with collisions occuring between objects in different belts too (f), at which point the swarm is largely randomized. \label{fig:BeltHeating}}
\end{figure*}

Now, when considering this question, an orbital crossing does not require the orbits to intersect exactly, they must merely come close enough for a collision. Because each element is spatially extended, the orbits have to miss by at least the diameter of the largest element. Effectively, it can be thought of as sweeping out a (generally eccentric) torus as it orbits its host. Several elements may be placed in the same orbit, maintaining a safe difference apart, where secular perturbations (section~\ref{sec:Perturb}) leave the orbits coincident, preventing collisions. To generalize the problem still further, we may imagine these elements on the ``same'' orbit receive small kicks over time, so that their orbits misalign, puffing out the torus (Figure~\ref{fig:BeltHeating}). Each torus is an orbital \emph{belt}, which serves as a unit of the swarm's structure. The inner radius of a belt $\rRin$ depends on the randomization of orbital velocities within it. A belt where the randomized velocities are so small that the inner radius is the radius of an individual element $\eRadius$ is referred to as ``sharp\editOne{''} in this work, whereas ``fuzzy'' in this work implies $\eRadius \ll \rRin \ll \rRout$, where $\rRout$ is the belt semi-major axis.

While there could be many elements in a belt, it is also evident that many belts are needed for some purposes. An occulter swarm presumably needs to ensure an occultation is seen from any angle, requiring a whole sweep of the range of inclinations; a Dyson swarm needs to collect luminosity emitted into all ecliptic latitudes. The belts are supposed to all fit within a rough shell or sphere with outer radius $\sRout$, so how many are allowed without crossing? 

In a swarm with $\rsN$ belts, a plausible ``packing factor'' is 
\begin{equation}
\rsPacking \equiv \rsN \rRin / \sRin .
\end{equation}
When the orbits are circular, belt intersections must occur if $\rsPacking \ga 1$, through the pigeonhole principle. Then all orbits that share a semimajor axis $\rRout$ are great circles on a sphere with that radius, and thus necessarily intersect. Non-intersecting circular belts thus each clear out a shell with thickness $\sim \rRin$, and the number of such shells that can fit in the swarm shell is $\sRin / \rRin$. Likewise, nested coplanar elliptical orbits intersect if $\rsPacking \ga 1$, with the nesting also imposing the same eccentricity and argument of pericenter.

This packing factor should also be appropriate for randomly distributed elliptical belts. Consider the midplane of a representative belt, which has a semimajor axis $\rSemimajor \approx \sRout$. The other belts, with nonzero inclinations, necessarily pierce the midplane twice. From their respective definitions, the midplane crossings are confined within an annulus centered on the host that has an area $4\pi\sRout\sRin$. We will assume that these are uniformly distributed; although high eccentricities imply many belt crossings close to the star, this concentration itself should increase the likelihood of an overlap. Overlap occurs whenever one of these crossings occurs within $\rRin$ of the belt, a target region with area $\sim 4 \pi \rSemimajor \rRin \sim 4 \pi \sRout \rRin$. Thus, by taking the area ratio, the expected number of overlaps occurring for this particular belt is $\sim 2 \rsN \rRin / \sRin \approx 2 \rsPacking$. Belt overlaps are common when $\rsPacking \ga 1$. In fact, the first overlap anywhere in the swarm occurs when $\rsPacking \sim \sqrt{\rRin/\sRin}$, because we need to repeat this analysis for every single belt, resulting in quadratic growth of belt crossings. Contrast that with nested circular belts, in which the first belt overlap anywhere in the swarm also occurs when $\rsPacking \sim 1$. This argument does not rule out the possibility that some ordered configuration (possibly with inclination and longitude of ascending node marching in step) allow for tighter packings.

A swarm with $\rsPacking \gg 1$ is therefore likely to be overpacked, with belt crossings. Overpacked swarms are potentially dangerous from a traffic control perspective. The phases of the orbits (mean anomalies) must be maintained, so that elements along belt slip through gaps in the elements along another. To maintain these phases in the faces of perturbations, all of the orbiting elements must maintain propulsion and navigation systems. Thus, overpacked swarms are unlikely to last long unless they are actively maintained -- they should not be expected to be long-lived passive technosignatures.

\subsection{The base configuration: how belts are likely to be arranged}
\label{sec:BaseConfig}
The collision rate is proportional to the differential velocity of elements with intersecting orbits. The differential velocity also sets the severity of the collision -- the faster the collision, the more debris produced, and the more of a threat the debris is to the remaining elements (Section~\ref{sec:Cascades}). As the velocities of elements within the belt are randomized by perturbations, the belts puff out (Figure~\ref{fig:BeltHeating}). Thus, adjacent belts can be expected to overlap first. If neighboring belts have high mutual inclination, then they begin to experience a barrage of high speed collisions as soon as they start to overlap. Low mutual inclinations means that the two streams of elements are moving almost cosynchronously, reducing the harms of collisions. To minimize the initial severity of collisions, adjacent belts should have nearly the same inclinations.

Circular orbits are also likely to be used. When there are other bodies in the system, elliptical orbits are well known to have drifting arguments of periastron and longitudes of ascending nodes, and variable inclinations and eccentricities, caused by secular perturbations like the Lidov-Kozai mechanism (see sections~\ref{sec:Perturb} and~\ref{sec:Radiative}). An elliptical orbit can be expected to sample the space all throughout a shell whose thickness spans the periastron and apoastron within a latitude range defined by the relative inclination \citep{Kessler81}. The higher the eccentricity, the thicker the shell segment. This essentially guarantees that belts overlap at some point. Furthermore, when collisions do happen, they have high velocity -- the velocity difference between two elements with significantly different arguments of periastron, or significantly different eccentricities, is of the same order of magnitude as the orbital velocity.

\begin{figure}
\centerline{\includegraphics[width=9cm]{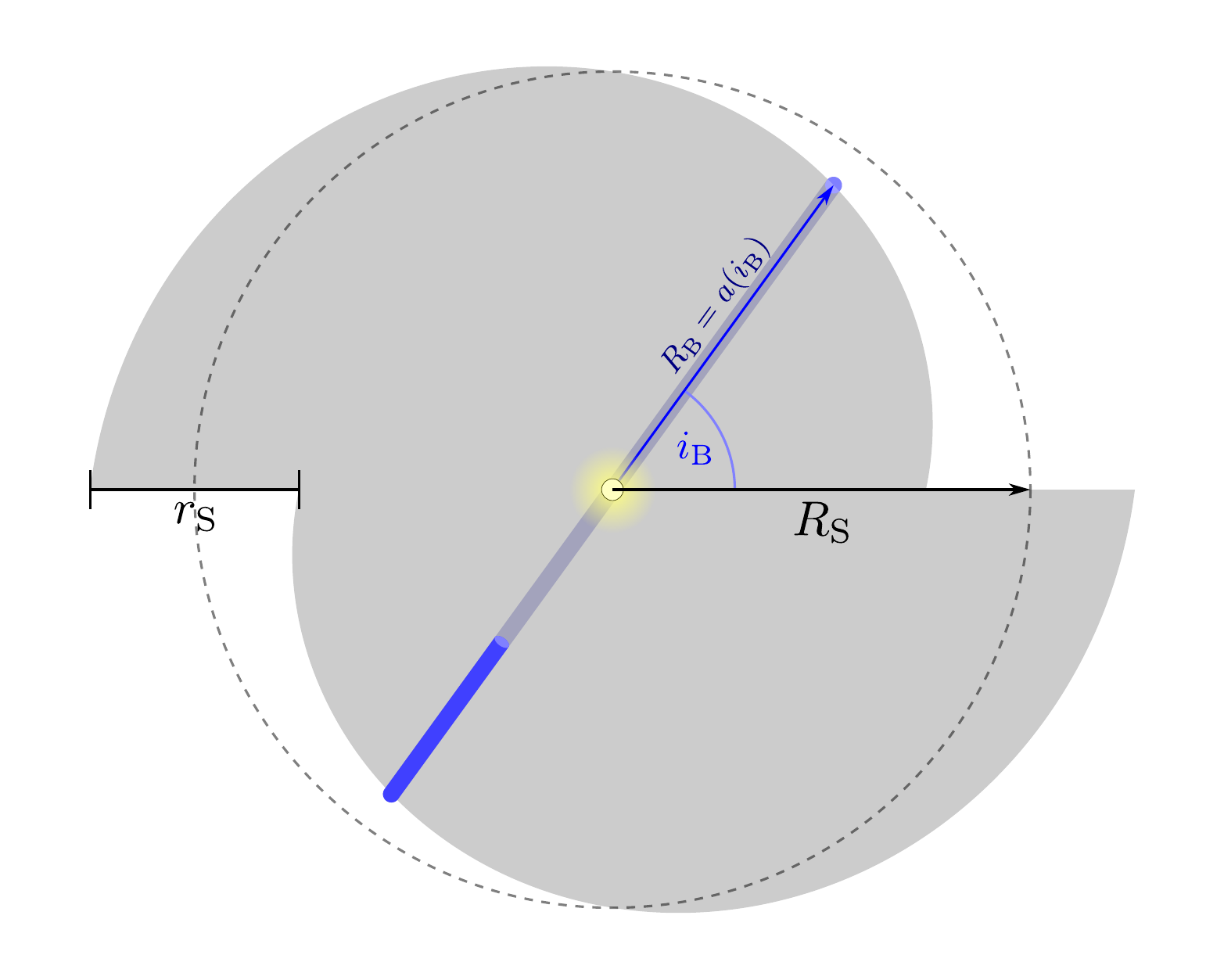}}
\figcaption{Sketch of the ``base configuration'' posited as a likely geometry for megaswarms, as viewed edge-on. The swarm includes circular belts (one example in blue) covering the full range of inclinations, but the semimajor axis gradually increases (or decreases) as it grows.\label{fig:BaseConfig}}
\end{figure}

As a result, an actual megaswarm built for longevity would use circular orbits organized on a gradient of inclination as one goes out from the star, increasing (or decreasing) steadily over a range of semimajor axes: the \emph{base configuration} (Figure~\ref{fig:BaseConfig}). This reduces the differential velocity of nearby belts. A corollary, however, is that most of the swarm shell's volume is unused if belt collisions are avoided. This has the effect of increasing the density within the occupied belts that much higher, shortening the intrabelt collision time.

\subsection{Collisions within belts}
\label{sec:intraBelt}
It makes sense that elements in a belt would originally be placed on exactly the same orbit, only at different phases. That way, long-term secular perturbations like the Lidov-Kozai effect influence the orbits in the same way (Section~\ref{sec:Perturb}), minimizing local velocity changes and helping to prevent collisions. Still, elements could acquire slight velocity offsets that desynchronize them and ultimately enable collisions, perhaps through the very act of station-keeping.

The elements continue to cluster around the original belt orbit, but at any time and phase of the orbit, the elements have a distribution of perturbations $\Delta \eVelVec$ relative to the circular orbit velocity $\rVCircVec$, with a total spread \editOne{$\rDispV$}. As a result, the elements have eccentricities that are slightly above zero and inclinations spread in a range around zero. Their orbits thread through a thin torus, with an inner radius $\rRin \sim \rRout (\rDispV / \rVCirc)$ and total volume $\rVolume \sim 2 \pi^2 \rRout \rRin^2 \sim 2 \pi \rRout^4 {\rDispV}^2 / (G_N \hM)$, if the velocity perturbation distribution has spread $\rDispV$ and is isotropic.\footnote{If the velocity distribution is anisotropic and has different dispersions in-plane and out-of-plane, the inner radii in the corresponding directions are scaled proportionally.}

In a sharp belt, the orbital torus is well within the tori swept out by the elements as they revolve around the host. The small offsets in the radial and out-of-plane directions are irrelevant for whether a collision happens, because they are smaller than the elements themselves. All that matters is whether the elements happen to have the same phase, at which point collision is inevitable. This is a one-dimensional problem, then; the collision time thus only depends on the linear density $\erLinDensity = \erN / (2 \pi \rRout)$ and the velocity spread (strictly, only the component parallel to the orbit):
\begin{equation}
\label{eqn:BeltTColSharp}
\rTCol \approx (\erLinDensity \rDispV)^{-1} \approx \frac{2 \pi \rRout}{\erN \rDispV} \approx \frac{\rPeriod}{\erN} \frac{\rRout}{\rRin} 
\end{equation}
for a circular belt with orbital period $\rPeriod$. Thus collisions never happen with zero velocity dispersion, and, assuming well-separated belts, the collision rate is fastest when the belt reaches the transition to being fuzzy ($\rRin \sim \eRadius$), because the linear density is unaffected by the dispersion. At this threshold, 
\begin{equation}
\rTCol \approx \frac{\rPeriod}{\erN} \frac{\rRout}{\eRadius} \approx \rPeriod \cdot \frac{23,000}{\erN} \cdot \left(\frac{\rRout}{\AU}\right) \left(\frac{\eRadius}{\REarth}\right)^{-1} .
\end{equation}
Within an occulter swarm, intrabelt collisions can be expected within tens of thousands of years. Other swarms may have smaller elements, but to achieve a high covering factor, many elements must exist within each belt ($\erN \gg 1$). Of course, the more tightly packed the elements are on the belt, the shorter this time becomes. The maximum packing occurs when the elements form a nearly continuous segmented ring, $\erN \approx 2 \pi \rRout / \eRadius$, at which point the collision time's minimum is $\rPeriod / (2 \pi)$. But the transition to fuzzy belts occurs for very small velocity dispersions,
\begin{equation}
\rDispV \approx \frac{2 \pi \eRadius}{\rPeriod} \approx 130\ \cm\,\sec^{-1} \left(\frac{\eRadius}{\REarth}\right)\left(\frac{\rPeriod}{\yr}\right)^{-1} .
\end{equation}
Collisions at these speeds would not harm solidly built elements, although they might snap off delicate structures. Instead, the objects would probably either bounce or just stick together. 

After this point, as long as the belts remain well-separated ($\rsPacking \ll 1$), further isotropized velocity dispersion actually prolongs the mean time between collisions because the torus cross-section puffs out in two dimensions. The volume of the torus is $\rVolume \approx \pi^2 \rRout^3 (\rDispV / \rVCirc)^2$ for an isotropized velocity distribution. Because the elements are randomly distributed in a three dimensional volume, the collision time is $\rVolume / (\erN \rDispV \eeAreaBAR)$:
\begin{equation}
\label{eqn:BeltTColFuzzy}
\rTCol \approx \frac{\rPeriod}{\erN} \frac{\rDispV}{\rVCirc} \left(\frac{\pi \rRout^2}{\eeAreaBAR}\right) \approx \rPeriod \frac{\rsPacking}{\sFcover} \left(\frac{\pi \eRadius^2}{4 \eeAreaBAR}\right)\left(\frac{\sRin}{\sRout}\right)
\end{equation}
for a typical belt in the swarm, noting that $\sFcover \approx \erN \rsN \eRadius^2/(4 \rRout^2)$.

As velocity dispersion increases, the belt packing $\rsPacking$ also increases, until it reaches $\sim 1$ and the belts overlap. At that point, we have to consider not only intrabelt collisions, but collisions with elements in other belts that cross the expanded tori, discussed in the next subsection. So, except when the velocity dispersion is essentially zero, the instantaneous collision time is shorter than the naive collisional time. Again, such collisions may be too slow to adversely affect the elements (though see section~\ref{sec:Thermalization}), but the speeds required for disruption are likely much smaller than the orbital speeds.

\subsection{When the belts overlap}
\label{sec:interBelt}

On the chaotic end, the collisional time reaches a plateau when the belts have randomized so much that they overlap, nearing the naive collisional time for the swarm.

We consider a swarm with $\rsN$ circular belts with an even spacing, $\Delta \rRout = \sRin / \rsN$, in radius (also the mean semimajor axis of elements within the belt). Each belt partially overlaps $\rrCovered \equiv \lfloor \rRin / \Delta \rRout \rfloor = \lfloor \rsPacking \rfloor$ belts interior to it, and an equal number exterior. For a random belt $j$, the collisional time is
\begin{equation}
\rsTCol(j) \approx \Biggl[\rTCol(j)^{-1} + \sum_{\substack{\Delta j = -\rrCovered\\ \Delta j \ne 0}}^{\rrCovered} \rrTCol(j, j + \Delta j)^{-1} \Biggr]^{-1} .
\end{equation} 
The terms include one using an intrabelt collisional time $\rTCol(j)$, which is the collisional time only including elements within the belt, and an interbelt collisional time, $\rrTCol(j, k)$, which is the collisional time of an element in belt $j$ due to collisions with elements in belt $k$ only. This interbelt collisional time is
\begin{equation}
\rrTCol(j, k) = [\erDensity(k) \eeArea \eeVRel(\BeltJMark, \BeltKMark) \rrOverlap(j, k)]^{-1},
\end{equation}
with $\eeVRel(\BeltJMark, \BeltKMark)$ being the typical relative speed of elements in belts $j$ and $k$ and an overlap fraction $\rrOverlap(j, k)$, which is the fraction of the volume of belt $\BeltKMark$ that coincides with belt $\BeltJMark$. When the density of elements in all the belts is roughly equal,
\begin{equation}
\rsTCol(j) \approx \left[\sum_{\Delta j = -\rrCovered}^{\rrCovered} \eeVRel(\BeltJMark, \BeltJDJMark) \rrOverlap(j, j+\Delta j) \erDensity \eeArea \right]^{-1} .
\end{equation}

First, what if the belts are coplanar (last panel of Figure~\ref{fig:BeltHeating})? The belts overlap each other at all longitudes, and because the belts are evenly spaced,
\begin{equation}
\rrOverlap(j, j+\Delta j) \approx \max\left(1 - \frac{|\Delta j| \cdot \Delta \rRout}{\rRin}, 0\right) .
\end{equation}
The relative velocities between elements in two belts has two components: the random velocities within the belts and the difference in circular orbital velocities between the belts. Now, the whole reason that belts have expanded enough to overlap is that the random velocities are so high that the orbits intersect. For most of the overlapping belts, then, $\eeVRel(\BeltJMark, \BeltJDJMark) \approx \rDispV$. Near the edges of the belt, the difference in circular orbital velocities is $\rRin \cdot d\rVCirc/d\rRout \approx \rRin \rVCirc/\rRout$, which is also $\sim \rDispV$. This can roughly double the relative speeds, but the overlap fraction is small near the edges, so we can ignore this effect at the level of approximation we are working. It follows that
\begin{equation}
\rsTCol \sim \left[\rsPacking \erDensity \eeArea \rDispV\right]^{-1} \approx \left(\frac{\rsN \eeArea \rVCirc}{2 \pi \sRout^2 \sRin}\right)^{-1} = \frac{\pi}{2} \frac{\rVCirc}{\sVCirc} \sTColNAIVE ,
\end{equation}
with $\sVCirc = \xVCirc(\sRout)$.

On the other hand, if the belts are mutually inclined, then they overlap only when one belt pierces through another. The longitude range over which belts with mutual inclination $\rrInclination$ overlap is $4\rRin/ (|\sin \rrInclination| \rRout)$, when $|\sin \rrInclination| \ga \rRin/\rRout$ (thus excluding the case when they are retrograde as well). This means the overlap fraction is 
\begin{multline}
\rrOverlap(j, j+\Delta j) \approx \max\left(1 - \frac{|\Delta j| \cdot \Delta \rRout}{\rRin}, 0\right)\\
\cdot  \frac{2 \rRin}{\pi |\sin \rrInclination(j, j+\Delta j)| \rRout} .
\end{multline}
Because one belt clears the other entirely for most of its circumference, the velocity difference caused by vertical motions is greater than the random velocity dispersion. It can be shown that inclination causes a relative velocity difference of
\begin{equation}
\eeVRel(\BeltJMark, \BeltKMark) \approx 2 \rVCirc |\sin(\rrInclination(j, k)/2)| .
\end{equation}
So, in this mutual inclination range
\begin{align}
\nonumber \rsTCol & \approx \left[\erDensity \eeArea \rDispV \rrCovered \Mean{\sec \frac{\rrInclination}{2}}\right]^{-1} \\
& \approx \frac{\pi}{2} \frac{\rVCirc}{\sVCirc} \frac{\sTColNAIVE}{\Mean{\sec \frac{\rrInclination}{2}}}.
\end{align}
For small mutual inclinations, $\sec (\rrInclination/2) \approx 1$, and even for perpendicular orbits, it is only $\sqrt{2}$, so unless the belts are retrograde, it can be ignored. Once again, the collision time is near the naive collisional time for the swarm. This is because the short time spent crossing the collision zone is approximately compensated by the faster relative speeds in that region.

\subsection{Confinement within wide bands: phase space and the robustness of the collisional time}
\label{sec:RobustCollisionalTime}
Most of the space in the belts model is wasted, even if they are packed tightly, because a single belt excludes an entire thin shell. What if instead the elements are confined to wide bands, wherein every element has the same semimajor axis but the inclinations have a wide range? It turns out this is no help for the collisional time. The key is that the velocity spread along the central plane is supposed to consist of the vertical motions out of the plane along inclined orbits. It amounts to a vertical velocity spread at the midplane of $2 \rVCirc \sin \rInclinationMax$. The density trades against the relative velocity spread. Whenever the inclination spread $\rInclinationMax$ is small, the area at the higher latitudes of the shell is empty, and wasted. 

This is a fundamental characteristic of the phase space for objects in circular orbits through a spherical potential. We want the swarm to fill as much physical volume within the shell, while minimizing the volume in velocity space. But consider the elements in a small neighborhood around a single point on the midplane at one instant. After one-quarter period, they all are moving in nearly the same direction, but they are spread to maximum extent, spanning a distance $\sim \rRout \rInclinationMax$. By Liouville's Theorem, the phase space volume is conserved over this evolution \citep{Binney87}. If the elements started out with a small velocity spread, then the phase space volume is also small, and the elements must have a high space density throughout their orbits. This is also true for elements elsewhere in the orbit, or that reach the midplane at different phases. One cannot fill the shell without having high-speed orbital crossings. That is why the naive collisional time is a robust result.

\section{The true speed of cascades in randomized swarms}
\label{sec:Cascades}
Once collisions begin occurring, a cascade develops -- large pieces are removed and replaced with showers of small pieces. These in turn create still further showers, and so on. Although the average cross section of the fragments goes down, the number of impactors increases rapidly; the pieces of debris are what truly destroy the original swarm.

In this section, I use simple numerical simulations to demonstrate just how quickly the cascade develops once it starts. Cascade simulations have been frequently carried out to investigate the process in Earth orbit. These typically include the spatial configuration of the debris cloud, at least a radial dependence from Earth, as it can develop at different rates in different regions \citep{Kessler78,Rossi94,Wang99,Liou04}. The spatial configuration of a megaswarm is not known -- the cascade may be developing in a thin belt or a thick shell for all we know -- and in any case a mere order-of-magnitude estimate will suffice. Thus, I will use a one-zone model, where the entire cascade is assumed to be homogeneous, along the lines of a particle-in-box model like \citet{Talent92} or \citet{Lewis09}. This could apply to the inner core of a belt \editOne{or in the thick of a fully randomized shell swarm}, for example; the assumptions may break down near the edges, but the results should still give a rough sense for how the cascade proceeds.

\subsection{The cascade equation}
\label{sec:CasadeEquation}
Even without any spatial dependence, and adopting a fixed velocity distribution, the elements differ in their size, which can be indexed by mass. The changing mass distribution of swarm elements governs the rate of fragment generation. The evolution of a distribution function of swarm elements occurs in a state space, and is analogous to the kinetic evolution of a gas in phase space. Thus, a Boltzmann-like equation governs the mass density distribution $\xMassSpectrumInline$ \citep[\editOne{cf.},][]{Wang99,Lewis09}. For our one-zone model of a rapid interplanetary cascade, \editOne{I use}
\begin{equation}
\label{eqn:CascadeBasic}
\frac{\partial}{\partial \TimeVar} \xMassSpectrum = \xInjection - \xGammaC \xMassSpectrum - \frac{d}{d\eMass}\left[\xMassRateE \xMassSpectrum\right] 
\end{equation}
\editOne{\citep[cf.][]{Ginzburg76}.}\footnote{\vphantom{T}\editOne{This equation is familiar from one-zone models of cosmic ray propagation, where new particles are injected by sources, or are destroyed or created in collisions. The mass of the fragment plays an analogous role to the energy or momentum of a cosmic ray.}} The approach of this section is to numerically solve this partial differential equation, although Monte Carlo methods provide another route \citep{Wang99}. 

The first term, $\xInjection$, is the injection rate of elements, and describes the mass rate distribution that fragments (including those that retain the majority of the element's mass) appear in the swarm. Next is the $\xGammaC$ term, the instantaneous rate of destruction of elements as a result of collisions. This can occur through \emph{catastrophic} shattering of elements when the impactor is big or fast enough, but even a non-catastrophic impact may gouge enough mass out that the element essentially jumps to a much lower mass, removing it from its current position in the mass distribution. 

The final term is an \emph{erosional} term that is not included in simulations of Earth orbit cascades. When a tiny grain hits a massive element at hypervelocity, it still gouges out a tiny crater and releases mass\footnote{Assuming we can neglect the element's gravity.}. In a discretized numerical setup, the mass difference from a single impact is negligible compared to the size of the mass bin, to seemingly no effect. However, a thick enough cascade releases a tremendous amounts of dust, producing enough grains that the aggregate mass loss can be significant over a time step. This advective term conserves number and accounts for this steady erosion of mass.\footnote{Since the maximum mass decrement from cratering ($\sim 10\%$; \citealt{Rossi94}) is of the same order as the mass resolution in my models ($10^{0.01} - 1$), I only include the catastrophic collisions in the destruction rate; all of the cratering or non-catastrophic losses are included in the erosion rate $\xMassRateEPRIME$, even if the largest of these should decrement the mass by a couple of mass bins.}

An important difference between the catastrophic and noncatastrophic collisions is that the rate of catastrophic destruction is set by the \emph{number} flux of elements above threshold, while erosional losses are set by the \emph{mass} flux of elements below threshold. As the cascade proceeds, mass is redistributed from larger pieces to smaller ones. This preserves the mass in the swarm, but vastly increases the number density. 

An excluded term that is usually found in Earth orbit simulations is a lifespan term. Various effects remove small enough grains rapidly from the system. In Earth orbit, this is primarily due to atmospheric drag \citep{Kessler78,Rossi05}, which can remove macroscopic fragments from \editOne{l}ow Earth \editOne{o}rbit much more quickly than the decades/centuries it takes for a cascade to develop. This quickly terminates the mass distribution and must be accounted for. But in interplanetary space, the range of masses spanned may be enormous -- from the width of planets on the high end to literally microscopic on the low end. Radiation pressure from the host star may \editOne{blow dust out}, but is a much less effective way to terminate the cascade (see discussion in section~\ref{sec:DustFate}). For this reason, I model a very large range of masses and neglect drag.

\subsection{Dimensionless variables for a family of solutions and the terms of the equation}
\label{sec:CascadeDimensionless}

Although there is a multidimensional parameter space of swarm configurations, include distance from the host, host mass, element size, and swarm density, the main controlling factor of equation~\ref{eqn:CascadeBasic} is the impact speed relative to the impact strength. Otherwise, the different evolution amounts to a rescaling of variables. Each variable $\xArbitrary$ is scaled to a characteristic value $\xArbitraryBAR$:
\begin{equation}
\xArbitraryPRIME = \xArbitrary / \xArbitraryBAR .
\end{equation}
This procedure is applied to element mass ($\xMassPRIME$), element area ($\xAreaPRIME$), number density of elements ($\xDensityPRIME$), and their relative velocities ($\xVPRIME$). A characteristic timescale follows:
\begin{equation}
\xTBAR \equiv (\xDensityBAR \xVBAR \xAreaBAR)^{-1} ,
\end{equation}
and writing the dimensionless equations in terms of $\xTPRIME = \xTime / \xTBAR = \xTime \cdot \xDensityBAR \xAreaBAR \xVBAR$ simplifies them considerably.

I consider all the original elements to have an identical mass $\eMassBAR$ and collisional cross-section $\eeAreaBAR$, with mean relative speeds $\Mean{\eeVRel}$ and local number density $\eDensityBAR$. When we set these equal to our scaling variables ($\xMassPRIME = \xMass/\eMassBAR$, $\xAreaPRIME = \xArea/\eeAreaBAR$, $\xDensityPRIME = \xDensity/\eDensityBAR$, $\xVPRIME = \xV / \Mean{\eeVRel}$), then the governing timescale becomes the naive collision time in the original environment (belt or swarm):
\begin{equation}
\xTPRIME = \xTime \cdot \eDensityBAR \eeAreaBAR \Mean{\eeVRel} = \xTime / \xTCollNAIVE .
\end{equation}
Finally, I scale the kinetic energy of elements relative to the kinetic energy needed to shatter one of the original elements:
\begin{equation}
\xKineticPRIME = \xKinetic / (\eMassBAR \eStrength)
\end{equation}
where $\eStrength$ is the characteristic impact strength of the elements (section~\ref{sec:ImpactPhysics}). The main parameter for the system of equations is
\begin{equation}
\xVPARAM = \Mean{\eeVRel} / \sqrt{\eStrength},
\end{equation}
which determines the amount and mass distribution of fragments produced by impacts. Thus, the dimensionless equations using these scales describe the progress of the cascade compared to the naive collision time, as a function of how severe the collisions are measured by speed.

\begin{widetext}
The terms in the cascade equation can be recast into these dimensionless forms as well. For the rate of catastrophic destruction,
\begin{equation}
\xGammaCPRIME (\eMassPRIME, \xTPRIME) \equiv \xGammaC \xTCollNAIVE = \int_0^{\infty} \int_{\xVCPRIME(\eMassPRIME, \xMassPRIME)}^{\infty} \xMassSpectrumPRIME(\xMassPRIME, \xTPRIME) \eeAreaPRIME(\eMassPRIME, \xMassPRIME) \PDF{\eeVRelPRIME}(\xVPRIME) \xVPRIME d\xVPRIME d\xMassPRIME .
\end{equation}
This integrates over $\PDF{\eeVRelPRIME}$, the probability distribution of (dimensionless) relative velocities $\eeVRelPRIME$, and the (dimensionless) masses of possible impactors, $\xMassPRIME$. For a fixed impactor mass and target mass, there is a velocity threshold $\xVCPRIME(\eMassPRIME, \xMassPRIME)$, above which the impact becomes catastrophic for the target with mass $\eMassPRIME$ (section~\ref{sec:ImpactPhysics}), which serves as a cutoff in the relative velocity integral. Finally, the (dimensionless) cross-section for collision is $\eeAreaPRIME$.

Erosional losses occur through collisions with impactors moving slower than the catastrophic velocity threshold:
\begin{equation}
\xMassRateEPRIME (\eMassPRIME, \xTPRIME) \equiv  \frac{\editOne{\xMassRateE \xTCollNAIVE}}{\eMassBAR} = \int_0^{\infty} \int_0^{\xVCPRIME(\eMassPRIME, \xMassPRIME)} \xMassSpectrumPRIME(\xMassPRIME, \xTPRIME) \dAggMassPRIME(\eMassPRIME; \xMassPRIME, \eeVRelPRIME) \eeAreaPRIME(\eMassPRIME, \xMassPRIME) \PDF{\eeVRelPRIME}(\xVPRIME) \xVPRIME d\xVPRIME d\xMassPRIME .
\end{equation}
The (dimensionless) mass of debris lost from the \emph{target} $\ElementMark$ due to the impact with the projectile of mass $\xMassPRIME$ is $\dAggMassPRIME(\eMassPRIME; \xMassPRIME, \eeVRelPRIME)$.  We can also define a ``rate of destruction'' set by non-catastrophic losses:
\begin{equation}
\xGammaEPRIME (\eMassPRIME, \xTPRIME) \equiv \frac{\xMassRateEPRIME (\eMassPRIME, \xTPRIME)}{\eMassPRIME} \editOne{= \frac{\xMassRateE(\eMass, \TimeVar)}{\eMass} \xTCollNAIVE},
\end{equation}
the fractional mass loss rate of an object.

Last is the injection term, in which pieces with mass $\eMassPRIME$ are generated by collisions of projectiles onto larger targets:
\begin{equation}
\xInjectionPRIME (\eMassPRIME, \xTPRIME) \equiv \xInjection \frac{\xTCollNAIVE \eMassBAR}{\xDensityPRIME} = \int_0^{\infty} \int_0^{\infty} \int_0^{\infty} \xMassSpectrumPRIME(\xMassPRIMEONE, \xTPRIME) \xMassSpectrumPRIME(\xMassPRIMETWO, \xTPRIME) \dMassSpectrumONEPRIME(\eMassPRIME; \xMassPRIMEONE; \xMassPRIMETWO, \xVPRIME) \eeAreaPRIME(\xMassPRIMEONE, \xMassPRIMETWO)  \PDF{\eeVRelPRIME}(\xVPRIME) \xVPRIME d\xVPRIME d\xMassPRIMETWO d\xMassPRIMEONE .
\end{equation}
Ejecta are the result of the collision of projectiles with (dimensionless) mass $\xMassPRIMETWO$ on targets with (dimensionless) mass $\xMassPRIMEONE$; the allowed range of both masses must be considered. The mean (dimensionless) mass distribution, $\dMassSpectrumONEPRIMEINLINE(\eMassPRIME; \xMassPRIMEONE; \xMassPRIMETWO, \xVPRIME)$, describes the number of pieces of debris with mass $\eMassPRIME$ ejected from the \emph{target} ($1$) when the projectile ($2$) hits it. Note it does not include the debris ejected from the projectile; that is included in the integrals when the values of $\xMassPRIMEONE$ and $\xMassPRIMETWO$ are interchanged.\footnote{When the masses of the projectile and the target are equal, then since the pair of masses is covered only once, the debris mass distribution should be doubled. However, the rate of collisions is then multiplied by $1/2$ to avoid double-counting the same collision. Thus, the integrand remains the same.} Additionally, it does matter if the debris results from catastrophic or erosional impacts. 

The dimensionless cascade equation is
\begin{equation}
\frac{\partial}{\partial \xTPRIME} \xMassSpectrumPRIME (\eMassPRIME, \xTPRIME) = \xInjectionPRIME(\eMassPRIME, \xTPRIME) - \xGammaCPRIME (\eMassPRIME, \xTPRIME) \xMassSpectrumPRIME (\eMassPRIME, \xTPRIME) - \frac{d}{d\xMassPRIME} \left[\xMassRateEPRIME (\eMassPRIME, \xTPRIME) \xMassSpectrumPRIME (\eMassPRIME, \xTPRIME)\right] .
\end{equation}
\end{widetext}

Table~\ref{table:CollisionTimes} may be used to convert the dimensionless quantities into physical values for habitable swarms in different host stars.

\subsection{The physics of collisions}
\label{sec:ImpactPhysics}

The production of debris and the destruction of elements is characterized by an impact strength, $\eStrength$, which is defined in terms of energy per mass. Consider a target at rest with mass $\xMassONE$ and a smaller projectile. When the total kinetic energy of the projectile is below the threshold energy $\xMassONE \eStrength$, the impact is non-catastrophic or ``cratering'' (erosional). A fraction of the projectile's kinetic energy is spent excavating into the target and launching material as ejecta; some more is spent on excavating or shattering the projectile itself; \editOne{some into the kinetic energy of the target;} the rest heating the bodies or damaging their internal structures without ejecting debris \editOne{\citep{Fujiwara89}}. Above that threshold energy, the impact is catastrophic, and the target is completely shattered, broken into many pieces.

The integrals also include cases where the target is smaller than the projectile. A naive application of this criterion would suggest that even a very slow impact could completely shatter a small target, just because the massive projectile contains so much kinetic energy. However, little of this kinetic energy actually reaches the target. \citet{Greenberg78}, a canonical work in Solar System impacts, suggests that, when the projectile is smaller than the target, about half of the  kinetic energy goes into each body, doubling the kinetic energy needed to shatter the target. For simplicity, I ignore this halving of the available kinetic energy, which is not a rigorous assumption and may not apply in all cases. Instead, I treat the kinetic energy of the projectile to be that of the smaller of the two in the other's frame. It follows that the critical relative velocity for shattering the target body $1$ is, in dimensionless terms, 
\begin{equation}
\xVCPRIME(\xMassPRIMEONE, \xMassPRIMETWO) = \frac{\sqrt{2}}{\tilde{\xV}} \max\left(\sqrt{\frac{\xMassPRIMEONE}{\xMassPRIMETWO}}, 1\right) .
\end{equation}
With this formulation, the smaller object is always shattered if the impact speed is greater than $\sqrt{2 \eStrength}$ ($\xVPRIME \xVPARAM \ge \sqrt{2}$). 

Models of the debris use the results of experiments where hypervelocity projectiles are shot into solid targets, supplemented by inferences from collisionally-evolved populations of minor Solar System bodies like asteroid families and cratering \citep{Greenberg78,Rossi94}. The total amount of debris\footnote{Excluding the mostly intact body when the collision is non-catastrophic.} depends on whether the body is shattered or not. From \citet{Rossi94}, the mass from the target body $1$ is 
\begin{equation}
\label{eqn:EjectaAggMass}
\dAggMassPRIME(\xMassPRIMEONE; \xMassPRIMETWO, \xVPRIME) = \begin{cases}
                                                            \displaystyle \frac{1}{20} \min(\xMassPRIMETWO, \xMassPRIMEONE) \cdot (\tilde{\xV} \xVPRIME)^2 & \text{(erosional)}\\
                                                            \xMassPRIMEONE                                                           & \text{(collisional)}
                                                            \end{cases} .
\end{equation}
Just below the threshold for shattering, 10\% of the target mass is excavated.\footnote{Equation~\ref{eqn:EjectaAggMass} implies that some mass is lost at very low speeds, but there is a threshold speed well below $\sqrt{\eStrength}$ at which both bodies simply bounce with no harm done \citep{Greenberg78}. I do not include this ``rebound'' regime in my models here, although this mechanism could result in thermalization.} The pieces in the ejecta cloud have a power-law distribution in mass:
\begin{multline}
\label{eqn:EjectaPowerLaw}
\dMassSpectrumONEPRIME(\eMassPRIME; \xMassPRIMEONE; \xMassPRIMETWO, \xVPRIME) = \dAggMassPRIME(\xMassPRIMEONE; \xMassPRIMETWO, \xVPRIME) \\
\cdot \begin{cases}
			\displaystyle (2 - q) \frac{{\eMassPRIME}^{-q}}{(\dMassMAXPRIME)^{2-q} - (\dMassMINPRIME)^{2-q}} & \text{if}~q \ne 2\\
			\displaystyle \frac{{\eMassPRIME}^{-2}}{\ln (\dMassMAXPRIME / \dMassMINPRIME)} & \text{if}~q = 2,
			\end{cases} .
\end{multline}
The minimum mass $\dMassMINPRIME$ of the pieces of ejecta is essentially $0$; in my models, it is arbitrarily set to be $10^{-25}$ for normalization purposes, below the smallest mass in the simulation.

As the impact becomes more violent, at higher speeds, the pieces generally become smaller. A maximum mass for the debris is imposed, with the largest possible piece of debris having a mass
\begin{equation}
\dMassMAXPRIME(\xMassPRIMEONE, \xMassPRIMETWO, \xVPRIME) = \begin{cases}
                                                                \frac{1}{2} \dAggMassPRIME                                                                  & \text{(erosional)}\\
                                                                \displaystyle \frac{1}{2} \xMassPRIMEONE \left(\frac{\xVPRIME}{\xVCPRIME}\right)^{-\xi} & \text{(catastrophic)}
                                                               \end{cases}
\end{equation}
where $\xi$ is an index that depends on the detailed physics of the collisions. The case for catastrophic collisions is taken directly from \citet{Rossi94}, who in turn base it on \citet{Fujiwara77}'s experiments shooting hypervelocity projectiles into rock. If this assumption holds, it can be shown that the secondaries created from a collision of two equally-sized primaries are never large enough to destroy the primaries themselves -- although the chunks excavated when the secondaries hit the primaries might \editOne{be}. The power law exponent also steepens at higher speeds:
\begin{equation}
q(\xMassPRIMEONE, \xMassPRIMETWO, \xVPRIME) = \begin{cases}
                                                  5/3                                                                                       & \text{(erosional)}\\
                                                  \displaystyle \frac{2 + \dMassMAXPRIME/\xMassPRIMEONE}{1 + \dMassMAXPRIME/\xMassPRIMETWO} & \text{(catastrophic)}
                                                  \end{cases},
\end{equation}
asymptoting at $2$ in the hypervelocity limit \editOne{\citep{Rossi94}}. \citet{Rossi94} use $q = 1.8$ and $\dMassMAXPRIME/\xMassPRIMEONE = 1/4$ for erosional impacts, and according to \citet{Greenberg78}, the power law exponent varies with severity of impact and energy transport within the target body when considering rocky materials; I adopt notional values of $q = 5/3$ and $\dMassMAXPRIME/\xMassPRIMEONE = 1/2$ to smoothly match up with collisional impacts.

The simple models I present here, like the \citet{Rossi94} model of a Kessler cascade, assume a single impact strength, but this is clearly inadequate. Even when looking at Solar System objects, there are definite divisions in the population as one moves from larger bodies held together by gravity and smaller bodies with internal strength. The elements of a megaswarm may lack this division, having a much lower surface density to maximize area covered. Even so, we can probably expect different regimes in the impact strength as the pieces are smashed further and further. Whether or not they behave like asteroids, which are both expected and inferred to have size-dependent strengths \citep{Davis94,Bottke05}, artificial objects are probably very heterogeneous. On the largest scale, entire modules of a satellite may be detached quite readily, and fragile parts like booms could easily be snapped off. Below that, the pieces could be broken into individual structural members like plates and fasteners, with more strength. We can imagine some materials fraying further into tiny fibers bound together, like carbon nanotubes or even textiles. Breaking these would require still harder impacts.

\subsection{Further considerations and methods}
\label{sec:CascadeMethod}

I adopt a planar geometry for the mass, each element having an area $\eAreaPRIME = \eMassPRIME / 4$. The collisional cross section is then $\eeAreaPRIME = (\sqrt{\eAreaONEPRIME} + \sqrt{\eAreaTWOPRIME})^2$ \citep[as in][and subsequent works]{Kessler78}, set equal to $1$ if $\eMassONEPRIME = \eMassTWOPRIME = 1$. This is the most likely result if we are imagining giant occulters or solar collectors that are relatively thin to conserve mass. Beyond a certain point, the ejecta should have a roughly spherical geometry, as the individual pieces become smaller than the original element thickness, with $\eeAreaPRIME \propto {\eMassPRIME}^{2/3}$. The trouble is that, very small ejecta collectively have a huge area; as mass is processed down to smaller and smaller sizes, the rate of impacts of the tiniest grains on merely small grains increases, further accelerating the cascade. Wild numerical instabilities result. As we are mainly interested in the destruction of the original members of the swarm and its largest pieces, I only consider planar geometry. Note that actual satellites and debris area-mass ratios deviate from planar geometry \citep{Kessler78,Badhwar89}.

For the relative velocity distribution, I adopt a very simple single-valued distribution,
\begin{equation}
\PDF{\eeVRel}(\xV) = \delta(\xV - \Mean{\eeVRel}) .
\end{equation}
This simplifies calculation of the ejecta products, although it may miss effects resulting from a small fraction of high velocity impacts on the tail.

When simulating cascades in Earth orbit, considering a relatively small range of masses is justified. Small particles may be removed by atmospheric drag, while there are clear upper limits to the size of objects we have placed in Earth orbits. Neither of these applies to the elements of a megaswarm. Radiation pressure forces may eject small grains orbiting stars, but depending on the stellar luminosity, they only operate faster than the cascade for literally microscopic particles (section~\ref{sec:DustFate}). On the large end, the original intact elements of the swarm may be kilometer scale habitats like space stations for inhabited Dyson swarms, or even planetary-scale screens in an occulter swarm. A kilometer scale solid sphere has $10^{27}$ times the mass of a micron-scale spherical grain, and a planet-sized sheet with surface density of $\sim 1\ \SigmaUnits$ can be several orders of magnitude more massive. Furthermore, for the highest relative velocities considered, $\xVPARAM = 10^5$, fragments with mass $\xMassPRIME \sim 10^{-10}$ can completely shatter the original elements with $\xMassPRIME = 1$, and likewise those fragments in turn can be shattered by grains with $\xMassPRIME \sim 10^{-20}$. I choose to simulate a very wide mass range $10^{-20} \le \xMassPRIME \le 1$, using a logarithmically sampled mass distribution with a resolution of $d\log_{10} \xMassPRIME = 0.01$. In some cases, the mass range may be truncated by dust ejection, in which case the cascade may proceed more slowly.

The integration of the partial differential cascade equation is accomplished using a simple explicit (Euler) method. This lets me use upwind differencing to prevent mass from ``leaking'' to $\eMassPRIME > 1$. I integrate for 100,000 time steps, each with $\Delta \xTPRIME = 10^{-3} \xVPARAM^{-1.5}$, which I find to be small enough to prevent numerical instabilities in the planar geometry.\footnote{For $\xVPARAM \sim 10 \endash 10^5$, $\Delta \xTPRIME$ remains in the range $\sim (1 \endash 3) \times 10^{-4} \xTCascPRIMEEST$ (equation~\ref{eqn:EstCascadeTime}).} \editOne{The integration is also stoppped if $\xTPRIME \ge 10^{2.5} \xTCascPRIMEEST$, with $\xTCascPRIMEEST$ defined in equation~\ref{eqn:EstCascadeTime}.}

\subsection{Numerical results}
\label{sec:CascadeResults}
	
There are several ways to define how long the cascade develops within the models. One aim is to consider how long the individual original elements last until they experience significant damage; we might then consider how long until the number of elements in the original mass bin falls by a factor $e$. This is fairly robust if the losses are all catastrophic, with the original elements always being reduced to a spray of tiny fragments, but if continuous erosional forces dominate, the times depend on the width of the mass bin. At the opposite extreme, we can consider how long until all fragments have been ground away to ``dust'' below the minimum considered size; but this depends on the lower threshold we consider for mass, the destruction of the smallest fragments by sub-threshold debris is not modelled, and the models do not include other forces like Poynting-Robertson drag that remove dust. A compromise is to consider either the number of or mass in fragments with $\eMassPRIME \ge 1/2$, at least half the original size.

\begin{figure}
\centerline{\includegraphics[width=8.5cm]{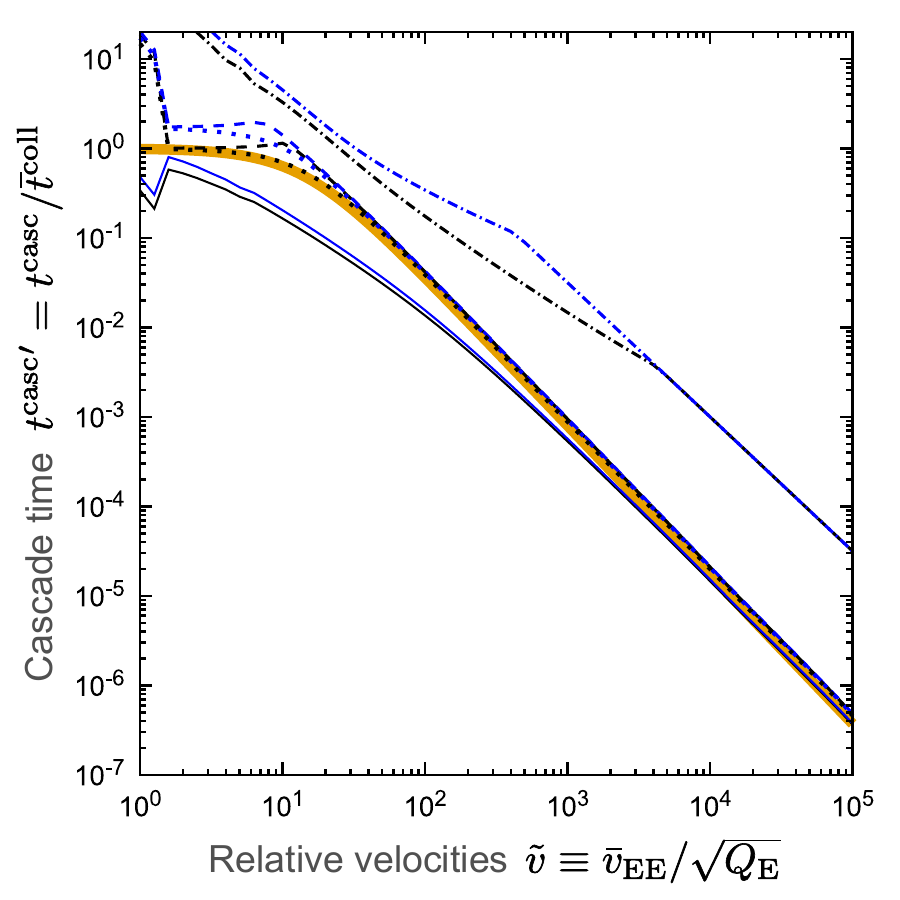}}
\figcaption{Computed time for destruction of original swarm according to the dimensionless cascade equations. Different line styles indicate different measures for the survival of the swarm: number of elements in the original mass bin (solid), number of elements (dashed) and mass in elements (dotted) at least half as massive as the original satellites, and total mass in macroscopic objects (dash-dotted). Black lines are half-lives and blue lines are $e$-folding times. An approximation to these times is provided by equation~\ref{eqn:EstCascadeTime} (\editOne{orange} solid). Refer to Table~\ref{table:CollisionTimes} to convert to physical times using the listed $\xTCollNAIVE$.\label{fig:CascadeTime}}
\end{figure}

However it is defined, the cascade time falls rapidly with increasing velocity, even relative to the naive collisional time $\xTCollNAIVE$ (Figure~\ref{fig:CascadeTime}). At high velocities, the various ways of computing the destruction of the original elements and the largest fragments converge to:
\begin{equation}
\label{eqn:EstCascadeTime}
\xTCascPRIMEEST = \frac{\xTCascEST}{\xTCollNAIVE} \equiv \frac{1}{1 + {\xVPARAM}^{5/3}/80} = \left[1 + \frac{1}{80}\left(\frac{\Mean{\eeVRel}}{\sqrt{\eStrength}}\right)^{5/3}\right]^{-1},
\end{equation}
(thick \editOne{orange} line), which amounts to a $\xTCascEST \propto {\xVPARAM}^{-8/3}$ dependence because $\xTCollNAIVE \propto \Mean{\eeVRel}^{-1}$. The estimated cascade times for habitable megaswarms around some archetypal stars is given in Table~\ref{table:CollisionTimes}.

Why is the cascade so rapid? At high speeds, the debris is extremely small, far too small to catastrophically damage the primary elements. These secondary fragments do inflict substantial erosional damage, digging out relatively large craters in the primaries as they impact. Some of the tertiary fragments from these impacts \emph{are} large enough to catastrophically destroy the primaries. Since catastrophic destruction is proportional to the number of elements instead of total mass, the conversion of primaries into these tertiaries greatly speeds up the destruction of the elements.

\begin{figure*}
\centerline{\includegraphics[width=18cm]{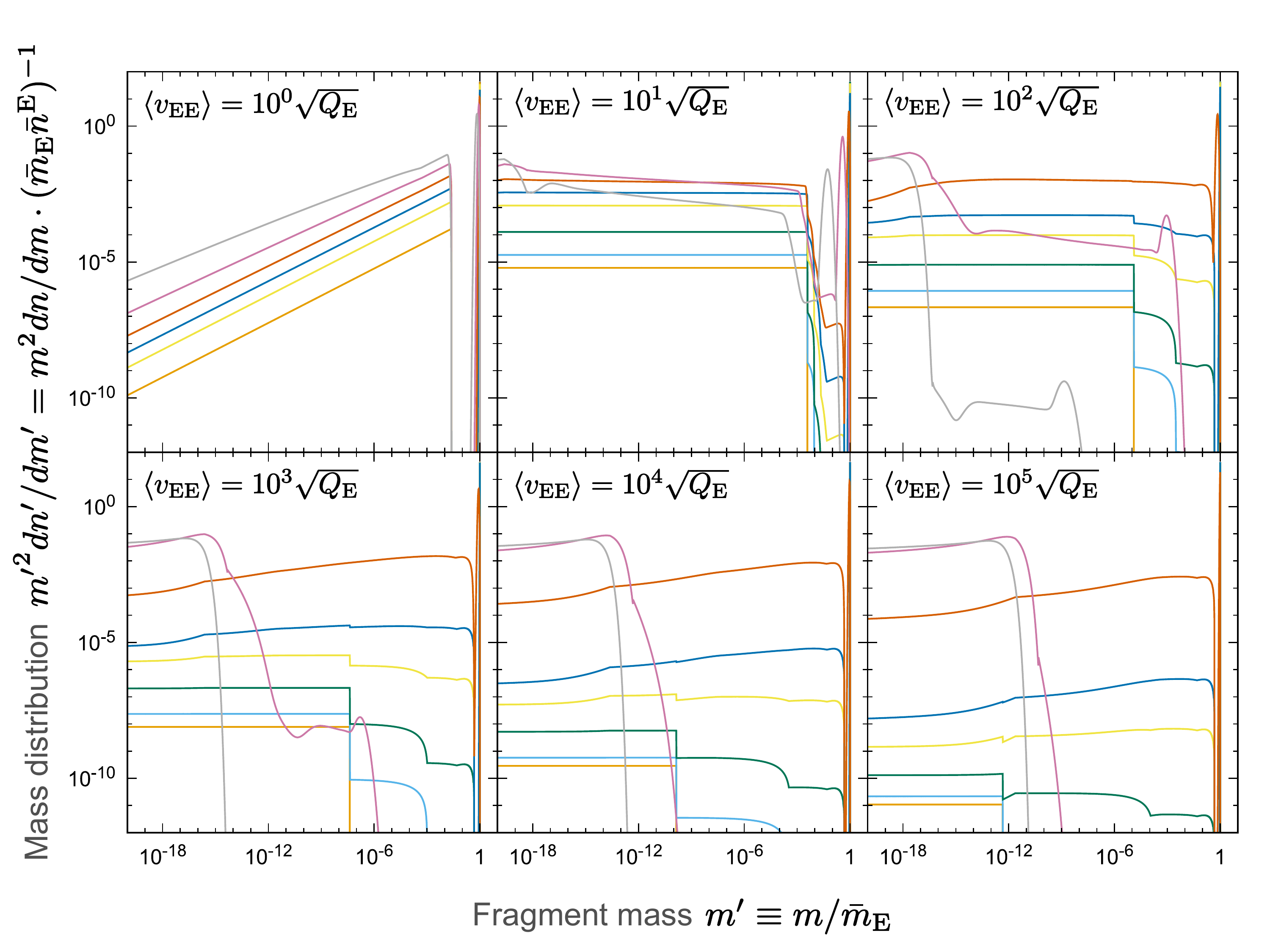}}
\figcaption{Mass distribution of the cascade according to the dimensionless cascade equations, as it evolves in cascades with different characteristic velocities. The distribution evolves from its initial shape (black), with all elements the same mass. After the first time step, the distribution follows the \editOne{light orange} curves. Later time steps shown correspond to $\xTPRIME / \xTCascPRIMEEST$ values of $10^{-3}$ (\editOne{light blue}), $10^{-2}$ (\editOne{dark green}), $10^{-1}$ (\editOne{yellow}), $10^{-0.5}$ (\editOne{dark blue}), $10^0$ (\editOne{dark orange}), $10^{+0.5}$ (\editOne{pink-purple}), and $10^{+1}$ (grey). \label{fig:CascadeSpectrum}}
\end{figure*}

This process is illustrated in Figure~\ref{fig:CascadeSpectrum}, which depicts the evolution of the mass distribution of objects in the swarm as velocity increases. The very first debris is from the collisions of two primary elements. These secondaries are all small, resulting in a significant gap in the mass distribution between the primaries and the debris (\editOne{orange} lines). The ejecta from the impacts of this first generation of the debris creates additional ``steps'' in the distribution from tertiaries and higher-order debris, as seen in the figure. At high enough speeds, this process manages to fill in the ``gap'' and results in a roughly power-law mass distribution extending nearly all the way to $\eMassBAR$. All along, the total amount of debris continues to build up with time. Eventually, the combined onslaught of the debris destroys the primaries, and the cascade proceeds to grind smaller and smaller elements away.

\begin{figure*}
\centerline{\includegraphics[width=18cm]{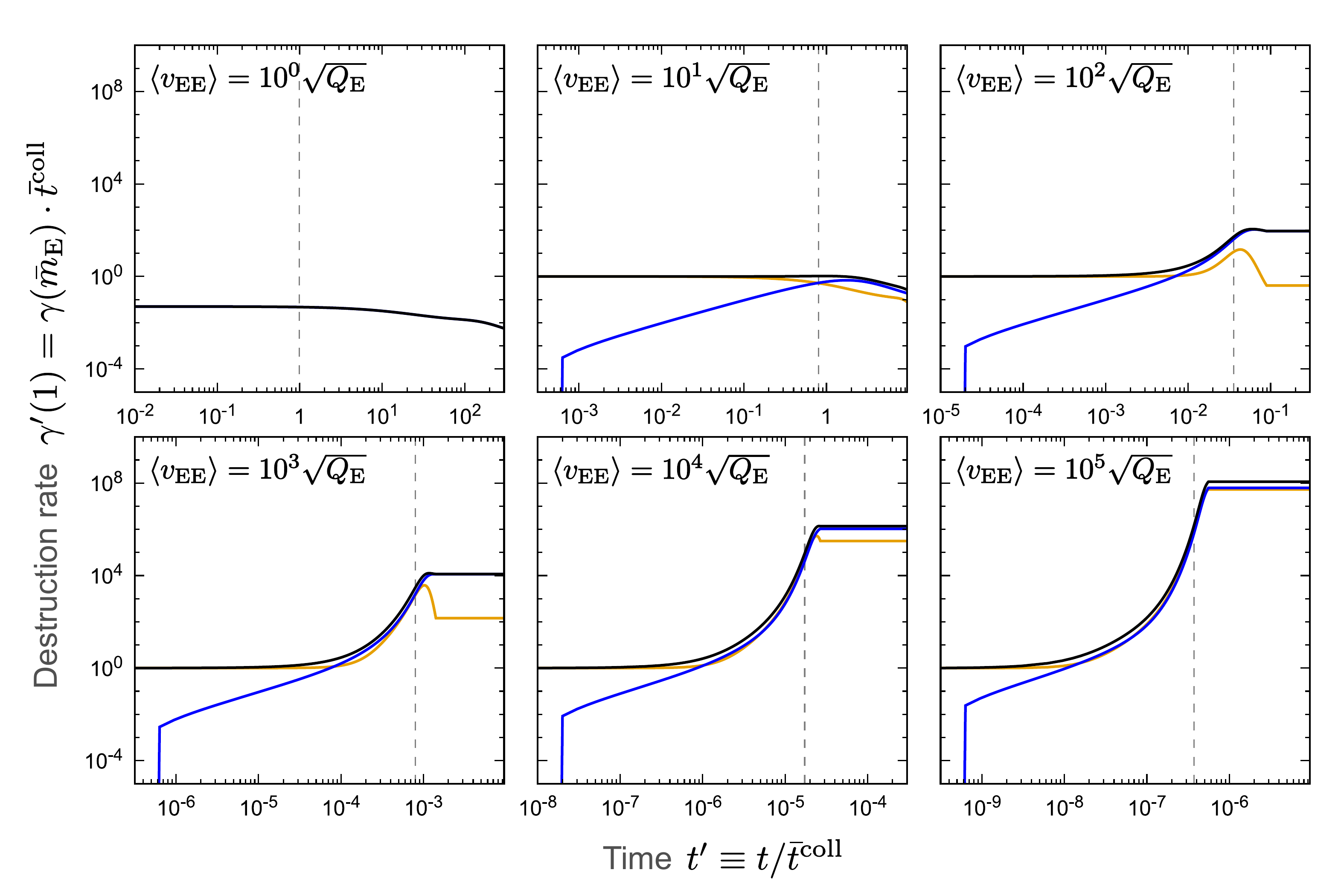}}
\figcaption{Effective rate of destruction that an element of the original mass ($\eMassPRIME = 1$, $\eMass = \eMassBAR$) would experience. At first the destruction is dominated by catastrophic shattering resulting from the original swarm members and the largest fragments (\editOne{orange}). As the collisional cascade proceeds, erosion from tiny fragments increases (blue) and dominates if the collisions are fast enough. The black line shows the total from catastrophic and non-catastrophic collisions. Dashed gray lines mark the estimated cascade time $\xTCascPRIMEEST$.\label{fig:CascadeRate}}
\end{figure*}

More insight into the destruction of the primaries is provided by Figure~\ref{fig:CascadeRate}, which presents the mean rate that a hypothetical object with $\eMassPRIME = 1$ would be destroyed or eroded. Initially, catastrophic destruction (\editOne{orange} lines) from collisions with other primaries is the main threat, except when $\xVPARAM \sim 1$. By itself, this would lead to an exponential fall over the collisional time. At early times, the debris builds up linearly, as does the erosional rate from the secondaries. The rate of increase is tremendous because of the extremely high speeds involved. At high $\xVPARAM$, when the losses from debris equal the losses from primary-primary collisions, the cascade takes off, with the losses increasing rapidly. 

This suggests a criterion for the time for a cascade to develop: $\xGammaEPRIME (\xTCascPRIME; \eMassPRIME = 1) \approx \xGammaCPRIME (\xTCascPRIME; \eMassPRIME = 1) \approx 1$. That occurs in a time of $\sim 80/{\xVPARAM}^2 \cdot \xTCollNAIVE$, a result of the debris being produced by collisions of the original elements (assuming $\xVPARAM \gg 1$). After that, the cascade then needs to accelerate by several orders of magnitude, which it does at a subexponential rate as the erosion-induced debris adds to the erosion (albeit with the debris itself being destroyed by other debris). An upper limit to the cascade time is found by ignoring this erosional debris-induced acceleration, only including the debris from the collisions of the original elements: $\xTCascPRIME \la 4\sqrt{10}/\xVPARAM$. Given that $\xTCollNAIVE$ itself is $\propto \xVPARAM^{-1}$, $\xTCascPRIME$ falls quadratically to cubically with characterisitc collision velocity.

Complete grinding to dust is slower. This seems to be because of a residue of objects with masses above threshold which is not efficiently destroyed, presumably because the debris from these objects is too small to be included in the model.

The case where $\xVPARAM \sim 1$ requires special consideration in my models, assuming the physics is accurate. Even a collision by two elements of equal size does not result in catastrophic destruction; instead $\sim 10\%$ is broken off from both. The total mass in the system remains the same, aside from losses as microscopic dust. Thus, the cascade is much slower and resembles an exponential fall over a duration $\xTPRIME \sim 10$ (Figure~\ref{fig:CascadeRate}).

\subsection{Implications from the strength of materials}
\label{sec:MatrialStrength}
Within the assumption of these models at least, increased speed above the threshold greatly increases the rate at which the swarm is ground away. But how strong should we expect the elements to be, and how do orbital speeds compare?

The canonical impact strength used for Earth satellites in most recent models is $4 \times 10^8\ \erg\,\gram^{-1}$ \citep{McKnight95}. Impact experiments have demonstrated catastrophic effects when the specific kinetic energy is twice this value \citep{Olivieri22}. It is well above the intrinsic impact strength of rocky materials in asteroids $\sim (1 \endash 3) \times 10^7\ \erg\,\gram^{-1}$ \citep{Davis94}. As noted above, the actual behavior is likely too complicated to model with a single number. Numerical simulations by \citet{Schimmerohn21} suggest that the modularity of satellites, with easily separated pieces, actually saves parts of the satellites by causing them to break off before they can shatter. 

But suppose we do assume that the elements of a Dyson swarm are roughly comparable to Earth satellites, rounding up to $10^9\ \erg\,\gram^{-1}$ to account for conservative future advances in material science. The characteristic impact speed for breakup is then
\begin{equation}
\label{eqn:StrengthVelocity}
\sqrt{\eStrength} = 0.32\ \kms \left(\frac{\eStrength}{10^9\ \erg\,\gram^{-1}}\right)^{1/2} .
\end{equation} 
This is much slower than orbital speeds in the Solar System, or Earth orbit, implying $\xVPRIME \sim 10 \endash 100$ and potentially very rapid cascades.

Of course, the builders presumably would know this, and if they want the swarm to survive as long as possible without constant guidance, the swarm would be arranged in a configuration that is likely denser but has smaller velocity differences (as in Section~\ref{sec:BaseConfig}). We might expect the actual cascade to begin with elements in circular bands, their eccentricities slowly increasing due to perturbations until they cross equation~\ref{eqn:StrengthVelocity}, and then proceeding at a rate limited by the perturbations.

One could question whether the impact strengths for Earth satellites is appropriate for an incredibly advanced Kardashev II \editOne{ETI}. Could they be using some kind of material that could survive impacts even at the orbital speeds within the Solar System? There are fundamental chemical limits, however. When the specific energies are sufficient to ionize individual atoms, probably no chemically or mechanically bonded atomic material could survive. The ionization energies of elements that can survive as solid materials in the expected conditions are at most $15\ \ev$, and the lightest such element likely to be used is carbon. This suggests a maximum strength of $1.2 \times 10^{12}\ \erg\,\gram^{-1}$, and a characteristic velocity of $11\ \kms$. A dynamically hot swarm of elements in orbit around the Sun in the inner Solar System would still suffer a cascade, if a slow one. Short of invoking some non-atomic material, collisional grinding is possible without careful planning for megaswarms around stars.

\subsection{The time of first impact: discreteness in finite swarms}
\label{sec:CascadeDiscreteness}
The cascade models assume that there is an infinite supply of elements to initiate the cascade. Thus, debris begins accumulating immediately. Actual swarms may have relatively few elements, perhaps about a hundred for occulter swarms. The cascade cannot begin until the first collision between the primary elements happens, releasing the first debris. In a randomized swarm, this is expected to occur around
\begin{equation}
\sTONE \equiv \frac{\sTColNAIVE}{\esN} \approx \frac{2 \pi \sRout^2 \sRin}{(\esN)^2 \eeArea \sDispV} .
\end{equation}
This sets a lower bound on the lifespan of a megaswarm. It is negligible for the archetypal Dyson swarm concept, but must be considered for sparse swarms, including minimal occulters.

An additional source of fragments is the impact of minor bodies on the elements. Stellar flybys can be expected to occasionally throw comets from the Oort cloud into the heart of the system, where they may hit elements at high speed. Even if the creators empty the entire Oort cloud, the swarm may experience impacts from interstellar objects \citep{Loeb23}.

\subsection{Low-velocity cascades: the thermalization threat}
\label{sec:Thermalization}
The base configuration described earlier suggests that local velocities start out small, and so the collisions themselves are slow, possibly well below the speeds needed for structural damage. Does this mean that cascades are not a threat?

Not necessarily. If the elements do bounce off one another, then there is a transfer of momentum changing both's orbits. The problem is that while velocity dispersion starts out small, the span of relative velocities across the swarm is high because of the need of a full range of inclinations. Over time, the bounces cause the velocities to randomize, and the swarm is thermalized.\footnote{\vphantom{T}\editOne{Thermalization, as used here, refers to the process whereby the organized gradients in velocities in a swarm are converted into a random distribution through collisions, without regard to the internal temperature of the elements and their fragments themselves. As noted in section~\ref{sec:Dissipation}, collisions are dissipative, however, and generate heat too, a separate effect than what is being discussed here.}} This is disastrous, as the hypervelocity spread in a randomized swarm annihilates the swarm in far less than the naive collisional time. Of course, what would actually happen is that the cascade begins long before full thermalization.

The thermalization process begins in earnest when the eccentricities of the swarm elements have been raised enough that their belts can overlap. Non-destructive collisions then happen, around one per element per collisional time $\sTColREAL$. If the swarm is densely populated but confined to a relatively few orbits, then the collisions do not occur until the orbits actually intersect; they have to find each other first, a process which is governed by a perturbation timescale for precession. Note this can be much shorter than a timescale for eccentricity evolution. If the swarm is very sparse, the orbits evolve quickly compared to the collisions, and thus are essentially randomized, so that $\sTColREAL \approx \sTColNAIVE$. 

At first, the velocity dispersion induced by collisions broadens each belt to only overlap with its nearest neighbors, because of the small velocity gradients in the base configuration. This alone would allow momentum to be transmitted from one end of the swarm to the other and eventually thermalize it, as neighboring belts influenced each other. That would be a relatively slow process: the effective sound speed is approximately given by the distance between belts over the collision time. It is likely, however, that the thermalization accelerates.

Locally thermalized belts should have fuzzy boundaries, and this provides a mechanism for rapid thermalization. If we consider a locally thermalized belt at time $\xTime$, the internal velocity dispersion is the result of the velocity gradient within the belt being converted into random motions: $\rDispV \sim d\xVCirc/d\SemimajorCore \cdot \rRin \sim \rVCirc \cdot \rRin/\sRin$. The relative velocity distribution will be centered at the circular velocity, but on the tails of the distribution are elements that delve further into the swarm where they come into contact with elements with high relative velocities. Those collisions transmit energy into the belt, heating it further, over the collision timescale $\sTColREAL$. This suggests that, over this timescale, the belt ``swallows'' some of the velocity gradient covered by elements on the velocity tails and thermalizes it. Since the length-scale of the fuzzy edges of the belt is $\sim \rRin \sim \rRout \rDispV/\rVCirc$,
\begin{equation}
\frac{d\rDispV}{d\TimeVar} \sim \frac{\rRin}{\sTColREAL} \frac{d\xVCirc(\rRout)}{d\rRout}  \sim \frac{\rDispV \rRout}{\sTColREAL \sRin},
\end{equation}
so that 
\begin{equation}
\rDispV(\TimeVar) \sim \exp\left(C \frac{\rRout}{\sRin} \frac{\TimeVar}{\sTColREAL}\right)
\end{equation}
for some constant $C$. If this argument holds, the entire swarm is thermalized (and destroyed) in perhaps tens of $\sTColREAL$.

To counteract thermalization, megaswarms may be forced to employ ``firebreaks'', large gaps between belts that must be overcome before momentum transfer can occur between them. These gaps shorten the collision time within belts, but this is worth it if the velocity spread within a belt is small enough to avoid damage during collisions.

The other possibility is that slow collisions are inelastic, and the elements stick. This does lead to evolution in the orbits, but thins the swarm as it proceeds, as elements aggregate. Additionally, the collisions tend to average the element velocities, making it less likely that some are bumped into the paths of others. Perhaps ETIs wishing to build long-lived megastructures choose to, for example, cover their satellites with hooks that snare with each other like velcro. Still, this also has the effect of thinning the swarm and reducing its effectiveness. An occulter swarm, for example, has fewer transits, leaving some inclinations with no transits. Also, even during inelastic collisions that are slow enough to prevent shattering of the swarm elements, there is still the possibility that small, loosely secured appendages break off during impacts. These form a population of particles that may themselves thermalize, turning them into high velocity missiles and dust that can erode the larger intact structures.

\subsection{The effects of dissipation on the cascade}
\label{sec:Dissipation}
\editOne{By assuming a fixed collision velocity, these models implicitly assume that no kinetic energy is dissipated in the collision. This, of course, is a gross simplification. While energy is dissipated in gases or plasmas by radiative mechanisms, in macroscopic swarms, dissipation results from the shattering of the structure of each body and internal heating \citep{Fujiwara89}, the latter of which may be radiated away long after the collision. Indeed, the impact strength in equation~\ref{eqn:EjectaAggMass} relates kinetic energy and debris mass, incorporating the physics of this energy loss mechanism.}

\editOne{The effects of dissipation are likely to be complex and geometry dependent. Consider a purely inelastic collision between two equal-mass elements on circular orbits with the same semimajor axis $\eSemimajor$ and mutual inclination $\eeInclination$. The resulting conglomerate is stationary in the center-of-mass frame. Thus, its inclination is averaged between the two original elements, and this averaging effect leads to a cooling of random velocities in a belt or swarm. The falling random velocities could slow down the cascade. One can also show that the fraction of orbital kinetic energy dissipated is $\sin^2(\eeInclination/2)$, which is also the resultant eccentricity; the new apocenter distance is the former $\eSemimajor$. Radial spreading results as the semimajor axis is slightly lowered. Over time, many collisions result in the swarm slowly shrinking as bulk kinetic energy is converted into heat.}

\editOne{Of course, the elements do not stick if the collision speeds are high enough; there is a spray of debris, each fragment of which can potentially cause more damage. Even if these fragments rapidly shed their random motions and settle into a dynamically cold disk, the mostly-intact original elements may still have relatively high inclinations with that belt. For our purposes, we are concerned most about whether the cooling suppresses random velocities before most of the original elements have been destroyed or eroded away. Drag effects require the element to interact with a mass about equal to itself. Meanwhile, the mass excavated by $\xMass$ through erosion is $\epsilon \xMass (\xDispV)^2 / (2 \eStrength)$, with $\epsilon = 0.1$ in the models. Dissipational effects greatly alter the cascade if the excavated mass is not much greater than the impactor mass, which happens when:
\begin{equation}
\xDispV \la \sqrt{2 \eStrength / \epsilon} \approx 1.4\ \kms\ \left(\frac{\eStrength}{10^9\ \erg\,\gram^{-1}}\right)^{1/2} \left(\frac{\epsilon}{0.1}\right)^{-1/2}.
\end{equation}
This condition is likely to hold within any isolated, narrow belts. If the belts overlap, however, and the swarm as a whole starts to thermalize, the random velocities approach $\sim \sVCirc$. Then, barring incredibly strong elements, impacts are so severe that objects in the swarm are obliterated before they can meaningfully damp their random motions and reduce $\eeVRel$.}

\editOne{The evolution of the mass spectrum as a whole is bound to be complicated, however, and a full simulation may be warranted.}

\subsection{The fate of the dust}
\label{sec:DustFate}

The cascade continues to grind down the fragments after the original elements are destroyed, converting more and more of it into tiny grains. For a brief time, the star may be shrouded in a veil of dust, a postmortem technosignature. The optical depth increases strongly because most of the mass of the original elements is buried in the interiors, already shadowed by their surfaces. Assuming there are $\gsN$ grains, all with the same radius $\gRadius$ and individual optical depth $\gOptdepth$, the optical depth of the swarm as a whole is $\sOptdepth \approx \gOptdepth \gsN (\gRadius/\sRout)^2$. Volume conservation gives us 
\begin{equation}
\gsN = \frac{\esN(0) \eVolume}{\gVolume} = \frac{3 \esN \eRadius^2 \eThickness}{4 \gRadius^3}
\end{equation}
for plate elements with an initial radius $\eRadius$ and thickness $\eThickness$.\footnote{For spherical elements, replace $(3/4) \eThickness$ with $\eRadius$.} Thus, if the initial covering factor of the swarm is $\sFcover(0)$,
\begin{equation}
\max \sOptdepth = \frac{3 \gOptdepth \sFcover(0) \eThickness}{4 \gRadius} = 7.5 \times 10^5 \gOptdepth \sFcover(0) \left(\frac{\eThickness}{\meter}\right) \left(\frac{\gRadius}{\um}\right)^{-1} .
\end{equation}
Starting instead with a minimal occulter swarm,
\begin{align}
\nonumber \max \sOptdepth & = \frac{3\pi \gOptdepth \eRadius^2 \eThickness}{16 \sRout \hR \gRadius} \\
& = 0.014 \gOptdepth \left(\frac{\eRadius}{\REarth}\right)^2 \left(\frac{\eThickness}{\meter}\right) \left[\left(\frac{\sRout}{\AU}\right) \left(\frac{\hR}{\Rsun}\right) \left(\frac{\gRadius}{\um}\right)\right]^{-1} .
\end{align}
A star that is $\sim 1\%$ ``too dim'' would not by itself stand out \citep{Zackrisson18}, but a brief opacity pulse would easily be detectable in modern photometry \citep[\editOne{cf.},][]{Montet16}, and the resultant infrared excess from reprocessed starlight is likewise visible \citep{Suazo24}. One caveat is that the absorption of starlight by dust grains falls below geometric when the grain is smaller than about one wavelength.

However, given the short timescales for a collisional cascade in a megaswarm, this opacity is likely to be very short-lived. If nothing else intervened, the grains would continue to grind down into plasma. In a fully thermalized swarm, even an individual carbon atom has a kinetic energy of $(1/2) m_C G_N \hM / \sRout = 56\ \eV\ (\hM/\Msun) (\sRout/\AU)^{-1}$, enough to ensure ionization during collisions unless the swarm starts out very far from the host star.

Radiation forces from the star are likely to blow out the dust grains unless the star is much dimmer than the sun, as is well-known from the dust supplied by minor body collisions in planetary systems (see also Section~\ref{sec:Radiative}). The key quantity is the ratio of radiation pressure forces to gravitational forces \editOne{\citep[][equation 19]{Burns79}},
\begin{align}
\label{eqn:GrainBeta}
\nonumber \gBetaRad & \approx \frac{\gOptdepth \hL \gArea/(4 \pi c \sRout^2)}{G_N \hM \gMass / \sRout^2} \approx \frac{3 \gOptdepth \hL}{16 \pi G_N c \hM \gDensity \gRadius} \\
&  \approx 0.19 \gOptdepth \left(\frac{\hL}{\Lsun}\right) \left(\frac{\hM}{\Msun}\right)^{-1} \left(\frac{\gDensity}{3\ \rhoUnits}\right)^{-1} \left(\frac{\gRadius}{\um}\right)^{-1}
\end{align}
Radiation pressure blows out dust when $\gBetaRad > 0.5$ if the dust start out in circular orbits, which should be true if the original elements have no radiation pressure support (\editOne{cf.}, Section~\ref{sec:Radiative}). The blowout occurs in about one orbital period, terminating the cascade and cutting off the mass spectrum.

Other effects that operate for larger grains include the forces of the solar wind and the radiation-induced Poynting-Robertson drag, which reduces the grain's semimajor axis on a timescale $\gTpr \approx (16\pi/3) \gDensity \gRadius \sRout c^3 / \hL$ \citep{Hughes18}. However, since $\gTpr$ has the same dependence on grain radius as the collisional timescale (since $\gsN \gArea \propto \gRadius$), Poynting-Robertson drag never overcomes the collisional cascade. This conclusion also holds for Solar System dust populations \citep{Wyatt02,Nesvorny10}.

\editOne{As for the remnant ion cloud around low lumiosity stars, a completely thermalized remant cloud in the habitable zone around a red dwarf would have a temperature around 
\begin{align}
\nonumber \xTemperature & \approx \frac{m_H \mu G_N \hM}{3 k_B \sRout} \left(\frac{\sDispV}{\sVCirc}\right)^2 \\
&  \approx 2 \times 10^6\ \Kelvin\,\left(\frac{\mu}{10}\right) \left(\frac{\hM}{0.2\ \Msun}\right)\left(\frac{\sRout}{0.032\ \AU}\right)^{-1} \left(\frac{\sDispV}{\sVCirc}\right),
\end{align}
where $\mu m_H$ is the mean mass of particles in the remnant plasma, and the other parameters are taken from Table~\ref{table:CollisionTimes} for an M dwarf. If we consider a Dyson swarm with mean surface density $\sSurfDensity$, the mean ion density is around
\begin{align}
\nonumber \xNDensityIon & \approx \frac{\sSurfDensity}{\mu_i m_H \sRin} \\
& \approx 1 \times 10^{13}\ \cm^{-3} \left(\frac{\sSurfDensity}{100\ \gram\,\cm^{-2}}\right) \left(\frac{\mu_i}{10}\right)^{-1} \left(\frac{\sRin}{0.032\ \AU}\right)^{-1}
\end{align}
with $\mu_i m_H$ being the mean mass of an ion. Both the temperature and density would be higher for a megaswarm in the habitable zone of a white dwarf. The plasma should thus be both very dense and highly ionized. If it is optically thin, it could cool very quickly if it contains metals like iron or silicon, which radiate efficiently in this temperature range \citep[see][Figure 34.3]{Draine11}. Ultimately, it might settle down into a circumstellar disk, and perhaps start recondensing into a thin dust belt and ultimately planetary material. Given the vast range of possible megaswarm parameters and resulting conditions, further study is required to positively identify any distinct signatures.}

As a result, we expect the ultimate fate of megaswarms is to disappear almost completely. Around faint hosts like red dwarfs, white dwarfs, and substellar objects, the swarm is reduced all the way to a dilute cloud of metal ions. Those around brighter hosts are broken into dust grains that are dispersed into the interstellar medium by radiation pressure, too dilute to detect in absorption or emission.

\section{Gravitational perturbations and megaswarms}
\label{sec:Perturb}
We have seen that even small amounts of randomness ensures collisions happen within the naive collision time, and that when the collisions are high speed, the cascade develops extremely quickly. This still leaves the possibility that elements are placed in completely ordered orbits, or are organized well enough that collisions are slow enough to be harmless. In this section, I consider gravitational perturbations that occur because the swarm is not an isolated system. These induce a minimum amount of randomness that can spell doom for a megaswarm.

In some cases, particularly occulter swarms located far from their hosts, the perturbation times may actually be much shorter than the collisional time. Depending on how quickly the cascade develops and if there is sufficient eccentricity pumping, these swarms might essentially always be randomized, at least locally. ETIs may not bother ensuring the elements are on different orbits because collisions are so rare and perturbations will disturb neat belts anyway, although they probably would still use something like the base configuration to slow down the resultant cascade.

\begin{figure}
\centerline{\includegraphics[width=8.5cm]{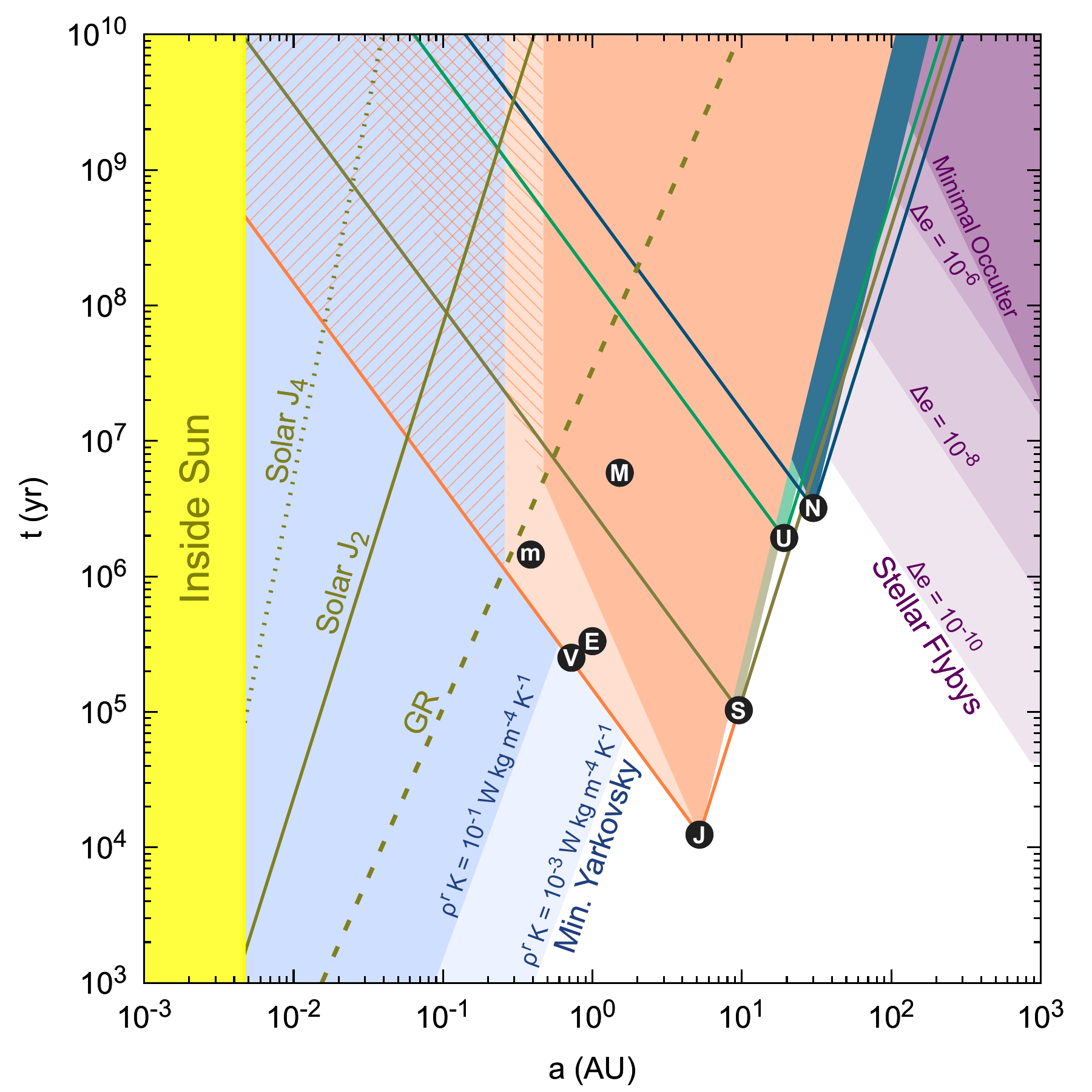}}
\figcaption{Characteristic perturbation times within the Solar System for an object orbiting at different semimajor axes. Shading indicates mechanisms that can pump eccentricity, in addition to precession. The quadrupole perturbation times from Venus (gray), Jupiter (pink), Saturn (gold), Uranus (green), and Neptune (teal) are shown, as well as the octupole perturbation times from Jupiter (dark pink). Additionally, the timescales for general relativistic precession and perturbations from solar oblateness are shown. Worst-case timescales for the Yarkovsky effect are shaded in blue, according to equation~\ref{eqn:tYarkovskyPlate}, for material properties appropriate for thin plates of solid metal (darker) and rock (lighter). The time it takes stellar flybys to pump eccentricities in the Solar neighborhood is shaded in purple, based on equation~\ref{eqn:FlybyEccentricity}.\label{fig:SolSys}}
\end{figure}

\subsection{How big of a perturbation can be tolerated?}
In an optimal configuration, elements with wildly differing inclinations are well-separated, probably by $\sim \sRin$, so as a rough estimate, the relative velocities between crossing elements can be estimated as that due to the spread in eccentricity. To order-of-magnitude, the relative velocities are sufficient to result in damaging collisions when
\begin{equation}
\rDispV \ga \rVCirc \rEccentricity \approx \sqrt{\eStrength},
\end{equation}
implying a critical eccentricity of around
\begin{equation}
\rEccentricity \approx \sqrt{\frac{\eStrength \eSemimajor}{G_N \hM}} \approx 0.011 \sqrt{\frac{\eStrength}{10^9\ \erg\ \gram^{-1}} \frac{\eSemimajor}{\AU} \left(\frac{\hM}{\Msun}\right)^{-1}}
\end{equation}
For elements orbiting near a solar analog's surface, this threshold shrinks to less than $10^{-3}$.

Thermalization may also trigger a cascade when belts begin to overlap (section~\ref{sec:Thermalization}). For that to happen, sufficient eccentricities are as small as
\begin{equation}
\rEccentricity \sim 1/\rsN .
\end{equation}

\subsection{Quadrupole and octupole-order secular perturbations from companion bodies}
\label{sec:HighOrderPerturb}
The discussion in Section~\ref{sec:CollisionOverview} assumed that the gravitational potential in which the swarm elements orbited was determined by a single host like a star, a monopole potential. But most stars have companions, be they stellar or planetary. On long time-scales, one can take a phase average of these bodies over the orbit, effectively replacing each body with a thin elliptical ring of varying density. This is the secular approximation \citep{Naoz16}. The presence of such a ring adds more terms to the gravitational potential, the lowest order nonzero terms being the quadrupole and octupole terms. 

\subsubsection{Perturbation timescales}
\label{sec:MultipolePerturbTime}
In general, the quadrupole and octupole-order terms may induce deviations in the inclination, eccentricity, longitude of ascending node, and argument of periastron. The semimajor axis $\eSemimajor$ is conserved, however. \citet{Naoz13} give equations for how these quantities evolve. For each perturber, I define a characteristic minimum perturbation timescale, the limit of the approximations as $\eSemimajor$ approaches the perturber's semimajor axis $\pSemimajor$:
\begin{equation}
\label{eqn:MultipolePerturbTimeMin}
\pTPerturbMin \approx \pPeriod \left(\frac{\hM}{\pM}\right) = \ePeriod \left(\frac{\pSemimajor}{\eSemimajor}\right)^{3/2} \left(\frac{\hM}{\pM}\right) . 
\end{equation}
Perturbations evolve more slowly far from the perturber, when $\eSemimajor \ll \pSemimajor$ or $\eSemimajor \gg \pSemimajor$. The timescale of quadrupole effects grows like $\epTQuad \propto \eSemimajor^{-3/2}$ as $\eSemimajor \to 0$ and $\eSemimajor^{7/2}$ as $\eSemimajor \to \infty$:
\begin{equation}
\label{eqn:tQuadrupole}
\epTQuad \approx \ePeriod \left(\frac{\hM}{\pM}\right) (1 - \eEccentricity^2)^{1/2} \cdot \begin{cases}
				\displaystyle (1 - \pEccentricity^2)^{3/2} \left(\frac{\pSemimajor}{\eSemimajor}\right)^3 \\
				\hfill\text{if}~\pSemimajor > \eSemimajor\\
				\displaystyle (1 - \eEccentricity^2)^{3/2} \left(\frac{\eSemimajor}{\pSemimajor}\right)^2 \\
				\hfill \text{if}~\pSemimajor < \eSemimajor
      \end{cases} 
\end{equation}
\editOne{\citep{Naoz13}.} To octupole order, these terms' dependencies steepen by another power:
\begin{equation}
\label{eqn:tOctupole}
\epTOct \approx \ePeriod (1 - \eEccentricity^2)^{1/2} \cdot \begin{cases}
				\displaystyle (1 - \pEccentricity^2)^{5/2} \left(\frac{\hM}{\pM}\right) \left(\frac{\pSemimajor}{\eSemimajor}\right)^4 \\
				\hfill \text{if}~\pSemimajor > \eSemimajor\\
				\displaystyle (1 - \eEccentricity^2)^{5/2} \left(\frac{\hM}{\pM}\right) \left(\frac{\eSemimajor}{\pSemimajor}\right)^3 \\
				\hfill \text{if}~\pSemimajor < \eSemimajor
      \end{cases} 
\end{equation}
\citep{Naoz13}. Here, the element is a test particle with negligible mass, and the perturber $\PerturbMark$ has a mass $\pM$ significantly less than the host as it orbits with a semimajor axis $\pSemimajor$ and eccentricity $\pEccentricity$.\footnote{In the notation of \citet{Naoz13}, $\epTQuad \equiv (\pi/8) |C_2 / G_1|^{-1}$ if the perturber is exterior to the element and $(15 \pi/32) |C_2/G_2|^{-1}$ if the perturber is interior; likewise $\epTOct \equiv (\pi/8)|C_3/G_1|^{-1}$ if the perturber is exterior and $(15\pi/32)|C_3/G_2|^{-1}$ if interior.}

\subsubsection{Perturbation evolution}
\label{sec:MultipolePerturbEvol}
The Lidov-Kozai effect is the most well\editOne{-}known example of quadrupole secular evolution, being implicated in all sorts of binary evolution phenomena \citep{Lidov62,Kozai62,Naoz16}. Suppose we have a perturbing object in a circular orbit outside a much smaller test particle, as is the case for a giant planet orbiting exterior to a swarm element. The out-of-plane angular momentum of the element is conserved, but inclination and eccentricity can be traded against each other. When the element's relative inclination is below a critical value, $39^{\circ}$, then a circular orbit remains circular.\footnote{Inclination is defined here with respect to the perturbing body, whose orbit defines the invariable plane when the element is a test particle. Note that precession of the ascending nodes can result in inclination changing when defined relative to another plane, like the midplane of the swarm.} An elliptical orbit precesses on the quadrupole timescale; additionally, the eccentricity may fluctuate slightly (as illustrated in Figure 4 of \citealt{Beauge06}). Beyond the critical value, however, an initially circular orbit at high inclination cycles into a highly eccentric orbit at low inclination and back, as the argument of periapse librates around $\pi/2$. This is highly relevant for megaswarm survival, because, unless elements at different inclinations start with extremely well separated semimajor axes, elements originally at high inclinations end up in the equatorial plane, colliding with the elements orbiting there  at high speed. This both ensures the rapid destruction of the megaswarm if a large enough perturbing body exists and also inhibits occultations and energy collection from the host's poles.

\begin{widetext}
The precession induced by the quadrupole perturbations is \editOne{\citep[][equations A29 and A30]{Naoz13}}
\begin{equation}
\label{eqn:eArgPerRate}
\eArgPerDOTQUAD = \frac{3\pi}{4\epTQuad} \cdot \begin{cases}
                                               4 \cos^2 \eInclination + (5 \cos(2\eArgPer) - 1)(1 - \eEccentricity^2 - \cos^2 \eInclination) & \text{if}~\pSemimajor > \eSemimajor\\
																							 2 + 3 \pEccentricity^2 + (5 \cos^2 \eInclination - 3)[1 + \frac{1}{2}\pEccentricity^2(3 - 5 \cos(2\pArgPer))] & \text{if}~\pSemimajor < \eSemimajor
                                               \end{cases} .
\end{equation}
There are also octupole terms, though they are of limited importance. This in itself can cause spaced-out eccentric orbits to cross. 

To quadrupole order, orbits exterior to the perturber experience no eccentricity or inclination evolution. Nor are circular orbits at low inclination affected by the Lidov-Kozai effect, which suggests that megaswarms can survive if the inclination range is within $\sim \pm 39^{\circ}$.\footnote{This type of partial Dyson sphere may be completely opaque near the equator and transparent near the poles. As space motions carried Earth in and out of the obscured inclination range, the star would appear to wink out and eventually back in when observed in the optical. Thus, if artificially ``disappearing'' stars are found, they may not represent a Dyson swarm being constructed, but the anisotropy of the swarm.} Octupole order perturbations provide a route for these quantities to change whenever the perturber itself has an eccentric orbit. Assuming the eccentricity is small, its rate of evolution is
\begin{multline}
\label{eqn:eccPEvolution}
\eEccentricityDOTOct \approx \frac{15\pi}{4\epTQuad} \eEccentricity \sin^2 \eInclination \sin(2\eArgPer) \cdot \IndicatorOf{\eSemimajor < \pSemimajor} + \frac{15 \pi}{32\epTOct} \pEccentricity [(5 \cos^2 \eInclination - 1) \sin \eArgPer \cos \pArgPer + (11 - 15 \cos^2 \eInclination) \cos \eInclination \cos \eArgPer \sin \pArgPer] \\
+ {\cal{O}}(\eEccentricity^2) ,
\end{multline}
where $\IndicatorOf{X}$ is $1$ if $X$ is true and $0$ if $X$ is false \editOne{\citep[][equations B10 and B11]{Naoz13}}. The inclination evolution is closely related:
\begin{equation}
\frac{d\cos\eInclination}{dt} = -\frac{\eEccentricity}{1 - \eEccentricity^2} (1 - \cos \eInclination) \eEccentricityDOT
\end{equation}
\citep{Naoz13}. 
The tempo of this higher-order eccentricity evolution is set by the rate of precession of the argument of periapse for both the element, from equation~\ref{eqn:eArgPerRate}, and the perturber, $\pArgPerDOTQUAD = 3\pi/(4\epTQuad) \cdot \cos \eInclination [2 + \EccentricityCore_1^2 (3 - 5 \cos(2\ArgPerCore_1))]$, where $\EccentricityCore_1$ and $\ArgPerCore_1$ are the eccentricity and argument of periastron of the element or the perturber, whichever is closer to the host star \citep{Naoz13}.\footnote{Since the element is a test particle, of course it cannot actually affect the perturbing body's orbit. The \emph{longitude of periapse} is the sum of $\pArgPer$ and $\pLonAsc$, with precisely cancelling precession rates. Inclination here is measured relative to the invariable plane; for a test element, the perturber necessarily has inclination 0, so its longitude of ascending node is arbitrary, with a non-zero precession rate that is still inconsequential.}
\end{widetext}

The exact solution is complicated and depends on several independent parameters. For our purposes, it suffices to note that $\eEccentricity$ oscillates, driven by the oscillations in the $\eArgPer$ terms over $\epTQuad$, because the cosines and sines of $\eArgPer$ and $\pArgPer$ are forced to remain between $-1$ and $1$ and fluctuate over $\epTQuad$. When we start with a circular orbit, or if the element orbits exterior to the perturber, only the octupole terms are nonzero. To order-of-magnitude, the amplitude of the eccentricity from the octupole terms is
\begin{equation}
\eEccentricityAMPLITUDE \sim \eEccentricityDOT \epTQuad \sim \pEccentricity \frac{\epTQuad}{\epTOct} \sim \pEccentricity \cdot \begin{cases}
                                                                                                                         \displaystyle \left(\frac{\pSemimajor}{\eSemimajor}\right) & \text{if}~\pSemimajor > \eSemimajor\\
                                                                                                                         \displaystyle \left(\frac{\eSemimajor}{\pSemimajor}\right) & \text{if}~\pSemimajor < \eSemimajor
                                                                                                                         \end{cases} 
\end{equation}
\citep[\editOne{cf.},][]{Ford00}.\footnote{The implied octupolar-driven eccentricity oscillation in Earth's orbit from Jupiter is $\sim 0.01$ over $\epTQuad \sim 10^5\ \yr$. Earth's eccentricity actually varies over a range of $0.06$ on long times, with large quasiperiodic variations over $\sim 100\ \kyr$. In fact, it is a complicated sum of many oscillations resulting from the influence of all the planets, though Jupiter is a governing contributor \citep{Laskar04,Spiegel10}.} Note that this amplitude does not depend on the mass of the perturber; only the timescale does.

Now, when the eccentricity is no longer non-zero, the quadrupole term in equation~\ref{eqn:eccPEvolution} can amplify these oscillations. We can evaluate the magnitude of that term at $\eEccentricityAMPLITUDE$ and compare to the octupole term, revealing that the key quantity is the inclination of the element. If the inclination is high, the quadrupole term is larger, allowing for the possibility of runaway eccentricity growth; if it is small, the quadrupole term is a minor correction. This is borne out by previous theoretical and numerical analyses \citep{Katz11,Lithwick11}. In fact, high-inclination orbits can ``flip'' between prograde and retrograde with maximum eccentricities near $1$, with the critical threshold dropping to about $\sim 60^{\circ}$ when $\epsilon \sim \epTQuad/\epTOct$ is greater than a few percent \citep{Lithwick11}. The timescale for this is roughly the geometric mean of $\epTQuad$ and $\epTOct$ \citep{Antognini15}. This effect is not relevant for our purposes, however; the minor eccentricity fluctuations at low inclinations are sufficient to trigger the cascade, destroying the swarm long before then.

\subsubsection{How common are fatal perturbers?}

\begin{deluxetable*}{llccccccc}
\tabletypesize{\footnotesize}
\tablecolumns{9}
\tablewidth{0pt}
\tablecaption{Some representative quadrupole and octupole perturbation times\label{table:HigherOrderPerturb}}
\tablehead{\colhead{Host} & \colhead{Perturber} & \colhead{$\pM/\hM$} & \colhead{$\pPeriod$} & \colhead{$\pSemimajor$} & \colhead{$\pEccentricity$} & \colhead{$\pTPerturbMin$} & \multicolumn{2}{c}{$\eSemimajor = 1\ \AU, \eEccentricity = 0$} \\ & & & & & & & \colhead{$\epTQuad$} & \colhead{$\epTOct$} \\& & & \colhead{($\yr$)} & \colhead{($\AU$)} & & \colhead{($\kyr$)} & \colhead{($\kyr$)} & \colhead{($\kyr$)}}
\startdata
Sun							 & Venus					& $2.45 \times 10^{-6}$ & $0.62$  & $0.72$    & $0.0067$ & $251$      & $780$             & $1100$\\
Sun              & Earth          & $3.00 \times 10^{-6}$ & $1.00$  & $1.00$    & $0.0167$ & $333$      & $330$             & $330$\\
Sun              & Jupiter        & $9.55 \times 10^{-4}$ & $11.9$  & $5.20$    & $0.0489$ & $12.4$     & $150$             & $760$\\
Sun              & Saturn         & $2.86 \times 10^{-4}$ & $29.7$  & $9.58$    & $0.0565$ & $104$      & $3100$            & $29000$\\
Sun              & Uranus         & $4.37 \times 10^{-5}$ & $84.1$  & $19.2$    & $0.0472$ & $1930$     & $160000$          & $3.1 \times 10^6$\\
Sun              & Neptune        & $5.15 \times 10^{-5}$ & $165$   & $30.1$    & $0.0087$ & $3200$     & $530000$          & $1.6 \times 10^7$\\
Earth            & Moon           & $0.0123$              & $0.075$ & $0.00257$ & $0.0549$ & $0.00608$  & $7.1 \times 10^6$ & $2.8 \times 10^9$\\
Jupiter          & Callisto       & $5.67 \times 10^{-5}$ & $0.046$ & $0.0126$  & $0.0074$ & $0.806$    & $3.6 \times 10^6$ & $2.9 \times 10^8$\\
Proxima Cen      & Proxima b      & $2.89 \times 10^{-5}$ & $0.031$ & $0.0486$  & $0$      & $1.062$    & $42000$           & $870000$\\
51 Peg           & 51 Peg b       & $4.13 \times 10^{-4}$ & $0.012$ & $0.053$   & $0$      & $0.028$    & $830$             & $16000$\\
$\alpha$ Cen A   & $\alpha$ Cen B & $0.843$               & $79.8$  & $23.3$    & $0.519$  & $0.0947$   & $9.0$             & $150$\\
Capella Aa       & Capella Ab     & $0.967$               & $0.285$ & $0.743$   & $0.0009$ & $0.000295$ & $0.0012$          & $0.0016$\\
61 Cyg A         & 61 Cyg B       & $0.877$               & $700$   & $85.9$    & $0.435$  & $0.798$    & $630$             & $44000$\\
Sgr A$^{\star}$  & S2             & $3.5 \times 10^{-6}$  & $16.1$  & $1030$    & $0.885$  & $4560$     & $1.5 \times 10^7$ & $3.4 \times 10^9$
\enddata
\tablerefs{Proxima system: \citet{SuarezMascareno20} and references therein, with $\sin \pInclination = 1$ assumed as the most conservative value; 51 Pegasi: \citet{Birkby17}; $\alpha$ Cen AB: \citet{Akeson21}; Capella A: \citet{Torres15}; 61 Cygni AB: stellar masses from \citet{Kervella08}, semimajor axis and eccentricity from \citet{Izmailov21}, with period derived; Sgr A$^{\star}$--S2: \citet{Abuter20}, using notional S2 mass of $15\ \Msun$ \citep[\editOne{cf.},][]{Chu18}}
\end{deluxetable*}

\begin{figure*}
\centerline{\includegraphics[width=8cm]{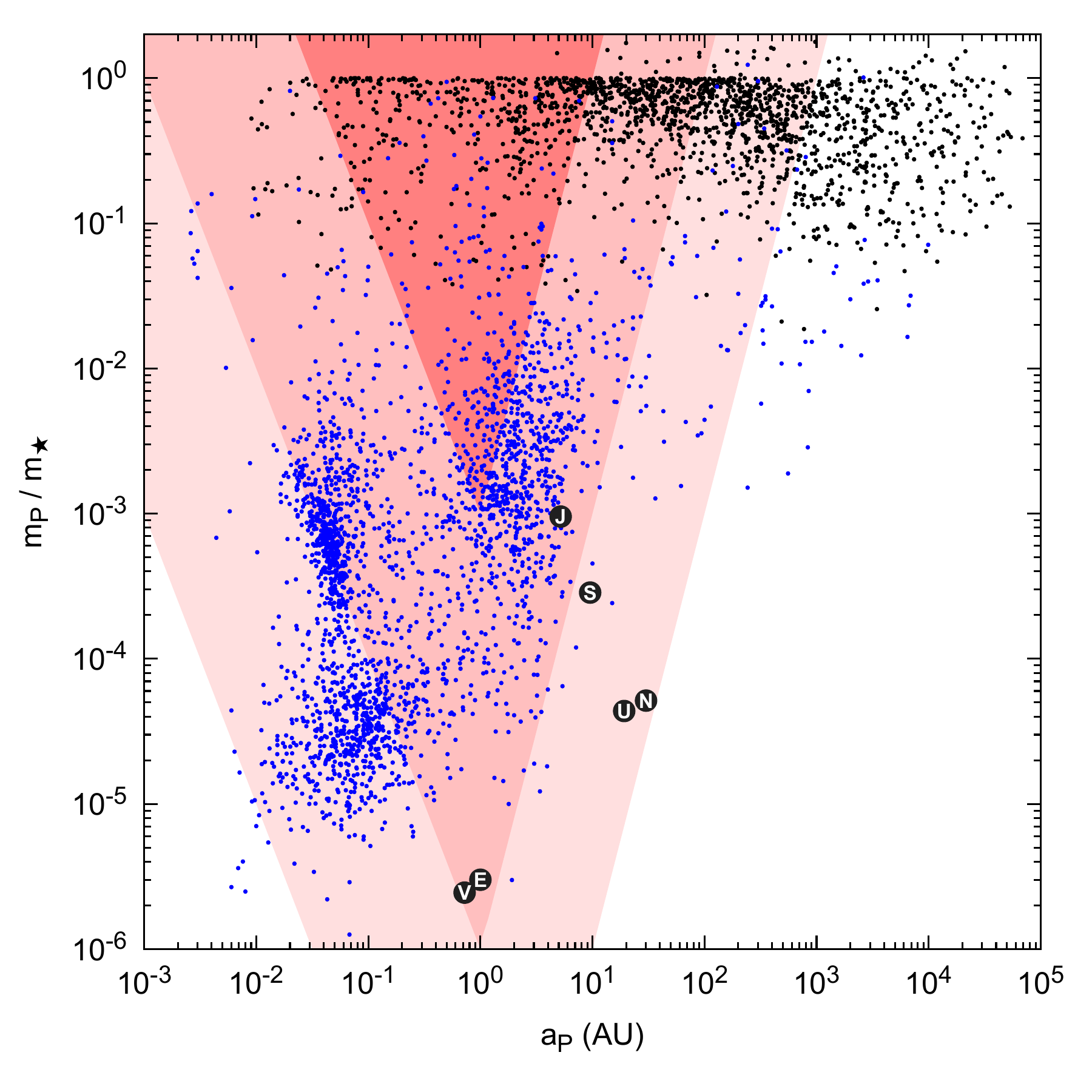}\includegraphics[width=8cm]{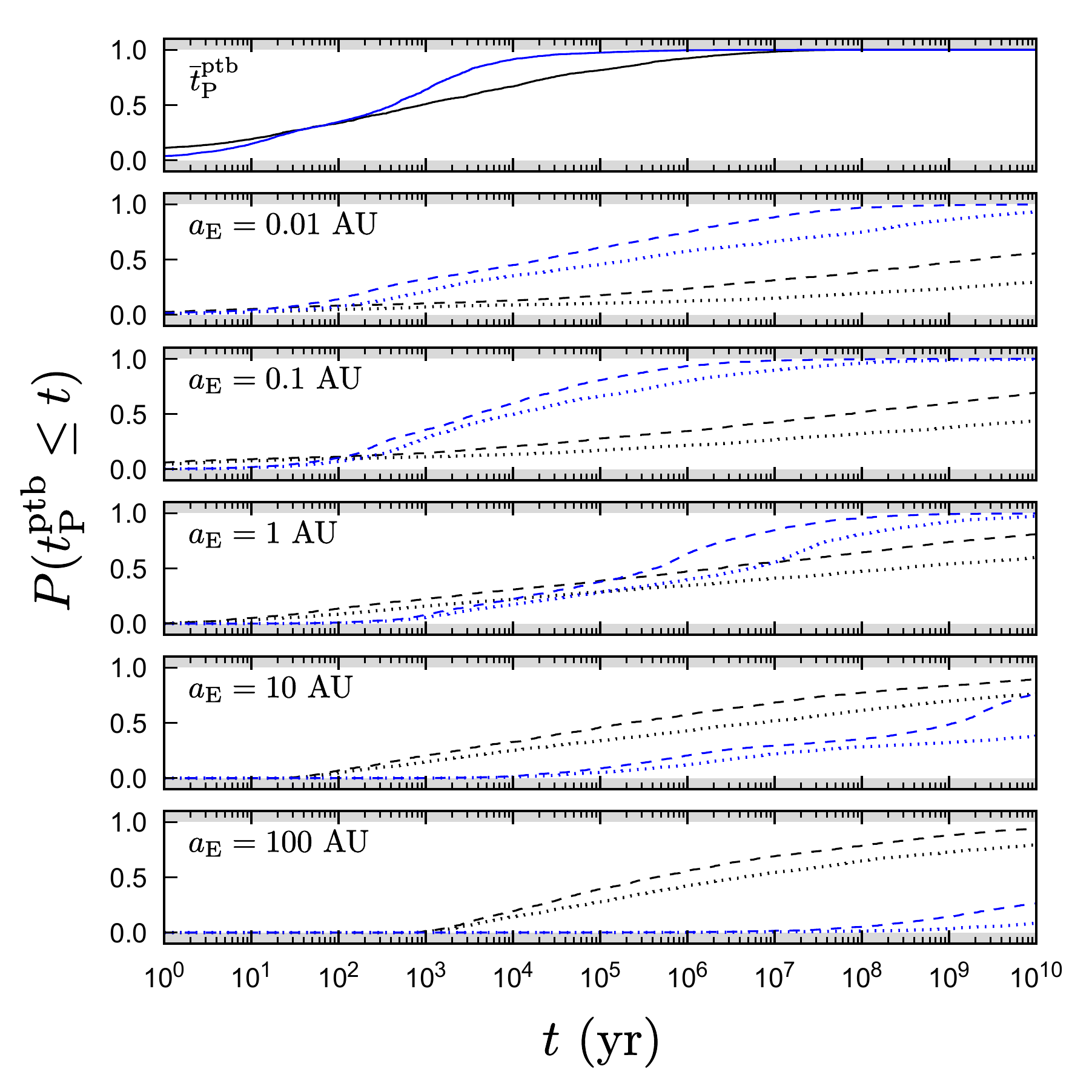}}
\figcaption{Illustrative figures for perturbing bodies orbiting stars, and their quadrupole and octupole effects. On left, a plot of semimajor axes and mass ratios for these companions. Black dots are stellar companions from \citet{Tokovinin14}; blue dots are exoplanets listed in {\tt exoplanet.eu}; also shown are the six largest Solar System planets. For comparison, pink-shaded regions indicate regions where a companion induces quadrupole perturbations within a given time on a test body in a circular orbit around a $1\ \Msun$ star at 1 AU (from darker to lighter, $10^3$, $10^6$, and $10^9\ \yr$). On right, the cumulative perturbation time distribution at various host distances from bodies in \citet{Tokovinin14} and {\tt exoplanet.eu}; $\pTPerturbMin$ is solid, while quadrupole times are dashed and octupole perturbation times are dotted. \label{fig:Perturbers}}
\end{figure*}

Table~\ref{table:HigherOrderPerturb}, listing perturbation timescales for some example systems, demonstrates the power of stellar companions to throw a megaswarm into chaos. Planets like Jupiter and Saturn would destabilize the swarm in a few million years or less. The problem gets much worse for close binary stars -- which is to say, a large fraction of stars with solar mass -- in which case destabilization can occur in millennia. 

In terms of the Solar System, the dominant source of quadrupole perturbations for most relevant semimajor axes is Jupiter (Figure~\ref{fig:SolSys}). Venus is very nearly as powerful interior to its orbit, as is Saturn exterior to its orbit, and Neptune dominates in trans-Neptunian space. 

For some perspective, I calculated quadrupole and octupole perturbation times for companions around nearby stars (Figure~\ref{fig:Perturbers}). Stellar companions are plotted according to the \citet{Tokovinin14} catalog, for stellar systems where the primary mass, secondary mass, and semimajor axis are listed (57\% of the full catalog, which may include several entries per system for hierarchical multiples). This selection does bias the sample by excluding close-in spectroscopic binaries and far-apart astrometric binaries without the required data. Among the selected systems, the median $\pTPerturbMin$ is 890 years. The median quadrupole time for a body orbiting at $1\ \AU$ is $2.9\ \Myr$, although the median octupole time is around $330\ \Myr$. About three-eights have $\ehTQuad(1\ \AU) \le 100\ \kyr$.

I also considered the exoplanets listed in the {\tt exoplanet.eu} archive \citep{Schneider11}. This is, of course, a highly biased sample because of the limitation of exoplanet detection methods: smaller and more distant planets are difficult to find. Nonetheless, a few basic conclusions are evident in Figure~\ref{fig:Perturbers}. A population of temperate Jupiters and super-Jupiters exists that would rapidly destabilize swarms at 1 AU. Quadrupole 1 AU perturbation times of around $10^3 \endash 10^5\ \yr$ are typical for these. Hot Jupiters induce quadrupole perturbations in a few million years, although these would not include eccentricity variations at 1 AU. Most of the remaining planets are hot Neptunes and super-Earths, with slower-acting perturbations.

Unless they are being actively stabilized or repaired, megaswarms in a large fraction of stellar systems will be destroyed quickly compared to geological timescales. 

\subsection{Perturbations close to the host}

\subsubsection{General relativistic precession}
General relativity ensures that the argument of periastron of an eccentric orbiting body evolves even if the central body is perfectly spherical. The Schwarzschild precession timescale is
\begin{align}
\nonumber \eTGR & = \ePeriod \frac{2 \pi}{\Delta \eLonAsc} \approx \frac{2\pi}{3} \frac{c^2 \eSemimajor^{5/2} (1 - \eEccentricity^2)}{(G_N \hM)^{3/2}} \\
& \approx 34\ \Myr\ \left(\frac{\hM}{\Msun}\right)^{-3/2} \left(\frac{\eSemimajor}{\AU}\right)^{5/2} (1 - \eEccentricity^2)
\end{align}
\citep[][\editOne{equation 58}]{Naoz16}. In a Schwarzchild metric, this precession is the only effect on the orbit from general relativity; there is no eccentricity evolution. The general relativistic precession can however suppress the Lidov-Kozai instability entirely if $\eTGR \ll \epTQuad$, although there is a window where $\eTGR \sim \epTQuad$ where it excites eccentricity instead \citep{Naoz13-PN}. The rapid Schwarzchild precession reduces the amplitude of the low-inclination octupole eccentricity variations proportionally to $\sim \eEccentricityDOT \eTGR$ (Section~\ref{sec:MultipolePerturbEvol}; \editOne{cf.}, \citealt{Holman97,Ford00}).

\subsubsection{Stellar oblateness}
Stars are not perfect spheres. Because they rotate, they are oblate. The nonsphericity introduces higher order multipole terms to the potential,
\begin{equation}
\Delta \hPotentialMultn(r, \phi) \sim \frac{G_{\editOne{N}} \hM}{r} \cdot \left(\frac{\hR}{r}\right)^n \hJMultn P_n(\cos \theta)
\end{equation}
for stellar latitude $\theta$, where $P_n$ is the $n$-th Legendre polynomial \citep[e.g.,][]{Mecheri21}. Oblateness is symmetric for the north and south hemispheres of the star, so only the even-order multipoles are included, starting with the quadrupole ($n = 2$) and hexadecapole ($n = 4$). The precession resultant from these terms inhibits Lidov-Kozai cycles, stabilizing eccentricity at high inclination.

Each multipole moment caused by the oblateness introduces a characteristic perturbation timescale, which is roughly
\begin{equation}
\ehTMultn \approx \frac{\ePeriod}{|\hJMultn|} \left(\frac{\eSemimajor}{\hR}\right)^n \approx \frac{2 \pi \eSemimajor^{3/2+n}}{\editOne{|\hJMultn|} \sqrt{G_N \hM} \hR^n}
\end{equation}
\citep[\editOne{cf.},][]{Iorio03}. Quadrupole order perturbations result in the precession of both the longitude of the ascending nodes and periastron, as well as advancement of the mean anomaly. The semimajor axis, eccentricity, and inclination remain constant, much like for an interior perturbing body \citep{Iorio11}. Hexadecapole order terms result in eccentricity and inclination changing as well. However, the hexadecapole eccentricity rate is itself dependent on the element's eccentricity, while limited by the quadrupolar precession of the pericenter over $\ehTQuad$:
\begin{equation}
\eEccentricityDOTHexd \approx \frac{15 \pi}{16 \ehTHexd} \frac{\eEccentricity}{(1 - \eEccentricity^2)^3} \sin(2\eArgPer) \sin^2 \eInclination (7 \sin^2 \eInclination - 6) ,
\end{equation}
and the inclination depends on eccentricity squared \citep[][\editOne{equation 3}]{Renzetti13}. This is in contrast to the eccentricity forcing by a perturbing body (equation~\ref{eqn:eccPEvolution}), which has terms that depend on the \emph{perturber's} eccentricity only. Even assuming some small deviation from $\eEccentricity = 0$, $\Delta \ln \eEccentricity \sim \ehTQuad/\ehTHexd \la 1$ (and $\ll 1$ for the Sun). Thus, the main effect of oblateness, like general relativity, is precession.

Late-type stars with convective envelopes spin slower as they age, as their stellar winds carry away angular momentum \citep{Kraft67}. Generally, the smaller and older the star, the slower its rotation, but stars more massive than the Sun spin quicker \citep[e.g.,][]{Barnes07,McQuillan14}. The Sun itself rotates slowly and has negligible oblateness. Measuring its multipole moments has been very difficult -- the differential rotation of the sun and its internal structure contribute to their values. But theory and observations suggest $\hJQuad \approx 2 \times 10^{-7}$ and $\hJHexd \approx -4 \times 10^{-9}$ \citep{Pitjeva05,Mecheri21}. So the associated timescales are
\begin{align}
\nonumber \ehTQuad & \approx 1.6\ \kyr \left(\frac{\eSemimajor}{\hR}\right)^{7/2} \left(\frac{|\hJQuad|}{2\times 10^{-7}}\right)^{-1} \left(\frac{\hDensity}{\rhosun}\right)^{-1/2}\\
          \ehTHexd & \approx 79\ \kyr \left(\frac{\eSemimajor}{\hR}\right)^{11/2} \left(\frac{|\hJHexd|}{4 \times 10^{-9}}\right)^{-1} \left(\frac{\hDensity}{\rhosun}\right)^{-1/2} .
\end{align}
As seen in Figure~\ref{fig:SolSys}, the quadrupole contribution from oblateness is expected to be greater than that of Jupiter and Venus out to $\sim 0.05\ \AU$, where the timescale is about ten million years, while the hexadecapole contribution operates much slower. 

Basically, solar oblateness should matter only in the zone of the Vulcanoids, hypothetical minor bodies orbiting the sun between $9$ and $45\ \Rsun$ out. In this region, collisional cascades are expected to be especially destructive because of the high orbital velocities \citep{Stern00}. Much smaller deviations in eccentricities are sufficient to ensure collisions damage the swarm elements.

Stars earlier than mid-F type do not have convective envelopes and so retain their primordial spins, approaching the critical velocity for breakup \citep{Kawaler87}. And in fact, many early-type stars are known to be oblate: A-type dwarfs like Vega and Altair and B-type dwarfs like Regulus have oblateness of order $\sim 10\%$ \citep{vanBelle01,Aufdenberg06,vanBelle12}. Although the multipole moments depend on internal structure, they should scale roughly with the ratio of the rotation speed to the orbital velocity squared \citep[e.g.,][]{Mecheri21}, and models indicate $\hJQuad \sim 0.005 \endash 0.02$ \citep{Zahn10}. 

Scaling to values appropriate for Vega ($\hREq = 2.8\ \Rsun$, $\hM = 2.1\ \Msun$): 
\begin{align}
\nonumber \ehTQuad & \approx 0.32\ \kyr \left(\frac{\eSemimajor}{10\ \hREq}\right)^{7/2} \left(\frac{|\hJQuad|}{0.01}\right)^{-1}\\
          \ehTHexd & \approx 3200\ \kyr \left(\frac{\eSemimajor}{10\ \hREq}\right)^{11/2} \left(\frac{|\hJHexd|}{10^{-4}}\right)^{-1} ,
\end{align}
for a distance where the temperature is $\sim 2,000 \endash 3,000\ \Kelvin$. Because the quadrupole timescale is much shorter than the general relativistic precession \citep[\editOne{cf.},][]{Iorio08}, oblateness is the main suppresant of the Lidov-Kozai effect.

Post-Newtonian effects resulting from stellar oblateness also result in evolution in the orbital eccentricity, but these are very slow for stars and, like the hexadecapolar term, $\eEccentricityDOT \propto \eEccentricity$, greatly limiting its ability to pump eccentricity \citep{Iorio19,Iorio24}.

\subsection{Internal secular perturbations}
The base configuration for a megaswarm -- a collection of rings in orbit around the star -- is also an instantiation of the ring model used to calculate the secular perturbations of a companion planet or star. Although the geometry is somewhat more complicated, since there are a large number of rings of differing inclinations, qualitatively we might model it as just a few rings, an inner and outer at zero inclination and a middle with $\rInclination \sim \pi/2$, similar to a system with a few mutually inclined planets. As a result, we expect the swarm mass to induce secular perturbations on itself, including the Lidov-Kozai effect. The timescale for these effects is roughly
\begin{align}
\nonumber \sTPerturbMin & \sim \ePeriod \frac{\hM}{\sM} \approx \frac{2\pi \sRout^{3/2} \hM^{1/2}}{G_N^{1/2} \hM} \\
& \approx 330\ \kyr\,\left(\frac{\sRout}{\AU}\right)^{3/2} \left(\frac{\hM}{\Msun}\right)^{1/2} \left(\frac{\sM}{\MEarth}\right)^{-1}
\end{align}
\editOne{from equations~\ref{eqn:MultipolePerturbTimeMin} and~\ref{eqn:tQuadrupole}.} The quadrupole Lidov-Kozai effect is likely to be most important in exciting eccentricity, as long as the inclination range of the swarm is sufficiently high. In contrast, since the belts are initially circular, they do not induce octupole-level pumping of the eccentricity until bent out of shape. Because the innermost and outermost belts in the swarm are not very widely separated in radius, however, even high-order terms in the multipole expansion may be important. Thus, an accurate treatment may need to include hexadecapole and even higher order contributions.

Full Dyson swarms are usually posited to have planet-scaled masses \citep{Dyson60}, so the rates should be similar to the effect of a perturbing planet:
\begin{align}
\nonumber \sM & \approx 4 \pi \sFcover \eSurfDensity \sRout^2 \\
& \approx 47\ \MEarth\, \sFcover \left(\frac{\eSurfDensity}{100\ \SigmaUnits}\right)\left(\frac{\sRout}{\AU}\right)^2,
\end{align}
for a swarm with a covering factor $\sFcover$ and surface density $\eSurfDensity$, which has a chosen value equivalent to a meter of water. Now, it is possible that the elements of megaswarms are more like dust, a distributed system of particles that extract power instead of consisting of macroscopic collectors or habitats. \citet{Wright20} pointed out there is a minimum surface density before radiation pressure blows out an element, assuming no countering thrust: $\eSurfDensityMIN \approx \hL/(4\pi G_N c \hM)$. Scaling to this value,
\begin{multline}
\nonumber \sTPerturbMin \approx \frac{2\pi c}{\hL} \sqrt{\frac{G_N \hM^3}{\sRout}} \frac{\eSurfDensityMIN}{\sFcover\eSurfDensity} \\
\approx 9.2\ \Gyr \left[\left(\frac{\sFcover \eSurfDensity}{\eSurfDensityMIN}\right) \left(\frac{\hL}{\Lsun}\right)\right]^{-1} \left(\frac{\hM}{\Msun}\right)^{3/2} \left(\frac{\sRout}{\AU}\right)^{-1/2} .
\end{multline}
So although swarms with macroscopic elements are self-destructing, these ``dust'' megaswarms are not much limited by self-driven Lidov-Kozai instabilities unless they are placed around very bright stars and thus require heavy ballast. Occulter swarms have so low a mass that the self-driven Lidov-Kozai effect is too slow to matter, unless the elements are kilometers thick.

As just another form of the Lidov-Kozai instability, general relativity and stellar oblateness can suppress these effects when those timescales are smaller than $\sTPerturbMin$.

\subsection{Stellar encounters}
Another source of perturbations is the collection of random tugs from passing stars as the system makes its way through its galaxy. This process is limited for megaswarms because they are compact compared to typical encounter distances. Almost all such encounters will be very distant, and thus gravitational focusing will be minimal. Over a time $\xTime$, the closest impact parameter of a star to the host is given by 
\begin{align}
\nonumber \pImpactParameter & \approx \frac{1}{\sqrt{\pi \Mean{\pV} \pNDensity \xTime}} \\
& \approx 2100\ \AU \left[\left(\frac{\Mean{\pV}}{30\ \kms}\right)\ \left(\frac{\pNDensity}{0.1\ \pc^{-3}}\right)\ \left(\frac{\xTime}{\Gyr}\right)\right]^{-1/2} ,
\end{align}
\editOne{where $\pNDensity$ is the number density of perturbing stars in the local neighborhood.}

Outside of star clusters, most stellar encounters are weak and distant -- much further from the host than any reasonably sized megaswarm. Even over geological timescales, the nearest expected flyby is extremely hyperbolic, with a pericenter speed much greater than the local escape velocity, and adiabatic, lasting much longer than the element orbital period \citep[\editOne{cf.},][]{Spurzem09}. Stellar flybys have different effects on circular orbits and elliptical orbits, with the possibility of decreasing the eccentricity for the latter, but as I approximate things, only the nearest encounter matters, its only considered effect to be to kick an initially circular orbit to a slightly elliptical one. The nearest encounter is described by the ``power-law'' regime of \citet{Heggie96}, with their equation 9 describing effects on circular orbits. In the limit of a distant encounter (with intruder eccentricity $\pEccentricity \to \infty$) and a test particle element, the eccentricity kick is\footnote{This is an octupole-level effect \citep{Heggie96}, evident from the cubic dependence on impact parameter.}
\begin{multline}
\eEccentricityKick \approx \frac{5}{4} \frac{\pM}{\hM} \left(\frac{\eSemimajor}{\pImpactParameter}\right)^3 \sqrt{\frac{G_{\editOne{N}} \hM}{\eSemimajor \pV^2}} \\
\cdot [\cos^2 \eInclination \sin^2 \pArgPer (1 - \sin^2 \eInclination (3 + \sin^2 \pArgPer))^2 \\
+ \cos^2 \pArgPer (1 - \sin^2 \eInclination (1 + \sin^2 \pArgPer))^2]^{1/2} .
\end{multline}
Although the kick depends on the geometry of the encounter, an approximate cross section $\pCrossSection$ is found by taking an \editOne{average over $\pArgPer$ and $\eInclination$ to find} a rough $\pImpactParameter$ \editOne{value} for each $\eEccentricityKick$:
\begin{equation}
\pCrossSection(\eEccentricityKick > \EccentricityCore) \approx \pi \pImpactParameter(\EccentricityCore)^2 \approx \pi \eSemimajor^2 \left[\frac{\editOne{0.614}}{\EccentricityCore} \frac{\pM}{\hM} \frac{\sqrt{G_{\editOne{N}} \hM/\eSemimajor}}{\pV}\right]^{2/3} .
\end{equation}
\begin{widetext}
After time $\xTime$, the element eccentricity will be around $\pCrossSection(\eEccentricity) \editOne{\pNDensity} \pV \xTime \sim 1$:
\begin{equation}
\label{eqn:FlybyEccentricity}
\eEccentricity(\xTime) \approx \editOne{3.4} \sqrt{G_{\editOne{N}} \hM \pV \editOne{(}\pNDensity \editOne{\xTime)}^3 \eSemimajor^5}\frac{\pM}{\hM}  \approx 2 \times 10^{-6} \left(\frac{\pM/\hM}{0.2}\right) \sqrt{\frac{\hM}{\Msun} \frac{\pV}{50\ \kms} \left(\frac{\editOne{\pNDensity}}{0.1\ \pc^{-3}} \frac{\xTime}{\Gyr}\right)^3 \left(\frac{\eSemimajor}{100\ \AU}\right)^5} .
\end{equation}
\end{widetext}

Stellar encounters are clearly very poor at raising eccentricity, although they may nonetheless be the only relevant mechanism for very wide megaswarms or in dense stellar clusters. A minimal occulter swarm in our Solar System, placed $100\ \AU$ out with about thirty thousand belts, would be safe for billions of years (Figure~\ref{fig:SolSys}). Dense swarms, designed for energy collection, however, may have many millions of belts; eccentricities $\ll 10^{-6}$ could be sufficient to result in orbit crossings, leading to swarm thermalization and eventual destruction in millions of years.\footnote{The assumptions used here eventually break down when (1) the time to reach an eccentricity threshold is shorter than than the time between individual stellar encounters ($\sim (\pNDensity)^{-1/3} \pV$); or (2), the ``power law'' regime of \citet{Heggie96} fails when $\pImpactParameter \sim \eSemimajor$.} 

Although this process is only conceivably important for dense or \editOne{far out} swarms, even with thermalization, its strength is its inevitability. Even if ETIs eliminate all other bodies in the stellar system (Section~\ref{sec:PlanetDestroyers}), short of ejecting the star out of the galaxy entirely, the kicks from stellar flybys are an intrinsic feature of the environment. Perhaps an extremely advanced Kardashev III metasociety could use stellar engines to manage the trajectory of every star in the region of interest, but this requires active maintenance -- it is not a way of preserving a passive technosignature without any effort.

\section{Radiation pressure: opportunity and menace}
\label{sec:Radiative}
\subsection{The radiative thrust timescale}
Radiation pressure from the host star influences the dynamics of subplanetary masses around a star. As discussed in Section~\ref{sec:DustFate}, it can blow small grains out of the system entirely, terminating collisional cascades; but it also has measurable effects on kilometer-scale asteroids in the Solar System \citep{Chesley03}. Radiative perturbations may have depleted minor body populations in the Solar System over its history \citep{Vokrouhlicky00,Bottke06}. 

Unlike gravitational perturbations, radiative accelerations are highly dependent on the geometry and composition of the objects they act upon. Additionally, radiative torques themselves can change the rotation rates of bodies. Rotation rate, density, shape, surface roughness, and thermal inertia all affect these forces \citep{Statler09}. Thus, radiative perturbations are characterized by dispersion -- objects on the same orbit get different kicks, randomizing their velocities, and leading to potential collisions (similar spreading is thought to occur in asteroid families; \citealt{Bottke06}). The problem may be particularly acute in Dyson swarms, where dispersion can lead to rapid thermalization and destruction of the swarm unless there are ``firebreaks'' (Section~\ref{sec:Thermalization}).

Only a limited amount of momentum is available from the host star, as parameterized by the ratio of radiative and gravitational forces,
\begin{align}
\label{eqn:BetaE}
\nonumber \eBetaRad & = \frac{\hL \eAreaRad}{4 \pi \ehDistance^2 c}\frac{\ehDistance^2}{G_N \hM \eMass} = \frac{\hL}{4 \pi G_N c \hM \eSigmaRad} \\
& = 7.65 \times 10^{-5} \left(\frac{\hL/\hM}{\Lsun/\Msun}\right) \left(\frac{\eSigmaRad}{\SigmaUnits}\right)^{-1},
\end{align}
where $\eAreaRad$ is the effective area of the element to incoming radiation from the host star, and the effective density column density $\eSigmaRad = \eMass / \eAreaRad$ \editOne{(see equation~\ref{eqn:GrainBeta} and \citet{Burns79})}. The effective area includes corrections for translucence. Now, since $\eBetaRad$ sets the ratio of accelerations, it also puts a bound on the timescale for radiative perturbations: 
\begin{align}
\label{eqn:TRadE}
\nonumber \eTRad & \ga \frac{\ePeriod}{2 \pi \eBetaRad} = \frac{2 G_N c \hM \eSigmaRad}{\hL} \ePeriod \\
& \ga 2080 \left(\frac{\hL/\hM}{\Lsun/\Msun}\right)^{-1} \left(\frac{\eSigmaRad}{\SigmaUnits}\right) \ePeriod .
\end{align}
How close the two timescales are depends on how the radiative forces are applied; anisotropy is needed for perturbations.

The relative strength of radiative perturbations has a clear dependence on type of star. Red giants and OB stars can have luminosity-to-mass ratios many hundreds times greater than the Sun, while the situation is reversed around red dwarfs and older white dwarfs (Table~\ref{table:CollisionTimes}).

\subsection{Radial forces: radiation as support and its precarity}
\subsubsection{The quasite option}
Radiative forces exerted in the radial direction on a swarm element can counter weight partially or entirely. Elements that use less material, perhaps for economical reasons, have a greater $\eBetaRad$ and thus this effect can be more important. In the limit that $\eBetaRad = 1$, an element becomes a statite, levitating in place without any transverse motion at all. This limit represents a minimum column density to a megaswarm \citep{Wright20}. For the $\eBetaRad < 1$ case, \citet{Kipping19} has proposed that ``quasites'' could use radiation pressure to partly levitate, reducing their orbital speed and prolonging their transits across stars. Effectively, the radiation pressure acts as if the mass of the host star has been reduced.

Radiation pressure support has an obvious advantage in inhibiting collisional cascades, in addition to being a necessary effect of reducing material usage. Because everything is moving more slowly, larger perturbations are required before collisions of elements become damaging. The radiation pressure support would have to be very strong for this to matter, though -- the resultant circular velocity is $\rVCircRad = \rVCirc \sqrt{1 - \eBetaRad}$. If the radiation pressure is half as strong as gravitational attraction ($\eBetaRad = 1/2$), for example, $\rVCircRad$ is decreased only by $1/\sqrt{2}$, and the eccentricity required for damage increased only by $\sqrt{2}$. 

\subsubsection{The unsteadiness of radiation support}
Yet, significant radiation pressure has disadvantages too, namely its unsteadiness. At the barest level, a non-spherical ``sail'' supporting the element can become titlted from normal orientation. This would reduce its lift and cause its orbit to shrink. Moreover, different sails could have different misalignments, inducing a velocity spread. There can be passive ways to stabilize the orientation (see \citealt{Lacki20-Lenses} for a hypothetical example), but there are other routes to degradation, including chipping of the reflective surfaces (e.g., from meteor impacts) or albedo evolution.

Nor are stars actually constant in luminosity. Late-type stars display magnetic activity, the visible manifestation being starspots and faculae. As the quasites pass over these features, they receive a small kick from the brightness fluctuation. Moreover, starspots and faculae are dynamic -- the kick received will depend on when in the \editOne{feature's} lifecycle the transit occurs. Kicks also arise from solar flares, although on the whole, these will be smaller because of their brevity. Over time, the randomness kicks will add up to a stochastic velocity spread even for elements in the same belt.

Stellar activity also has longer term effects. With the Sun, there is a well-known $0.1\%$ luminosity variability associated with the eleven year solar activity cycle, with times of high activity brightening the star \citep[e.g.,][]{Lean91}. Although it may seem like this would affect all elements equally, sunspots are concentrated on certain low- to mid-latitude ranges, while faculae may be found at the poles \citep[see][]{Makarov89,Solanki03}. During periods of high activity, elements at different inclinations can experience different radiative lifts on average, although admittedly the effect is small.

Other types of stars have more pronounced variability, like long-period variables and accretion-powered interacting binaries. Many of these are very luminous. Scrounging enough material to build a megaswarm at a distance where it would not sublimate could be difficult, suggesting that the elements would be lightweight and thus easily affected by radiation pressure variability \citep[as noted in][]{Lacki20-Lenses}.

\subsubsection{Secular luminosity increases and radiative blowout}
Even ignoring these effects, stars display secular evolution. Main sequence stars brighten as they age. Unless the elements adjust their properties to match, it will lead to the elements slowly being pushed further from the star. This reduces the megaswarm's optical depth, although the effects are important mainly for statites, which require a near-perfect balance of forces. They are eventually pushed out of the system entirely.

For a quasite initially on a circular orbit, or an initially stationary statite, the distance evolves as\footnote{This can be derived using the effective (mass-normalized) Lagrangian $\mathcal{L}_{\ElementMark} = (1/2) (\dot{\ehDistance}^2 + \ehDistance^2 \dot{\theta}_{\ElementMark}^2) + G_N \hM (1 - \eBetaRad(\TimeVar))/\ehDistance$.}
\begin{equation}
\label{eqn:eBlowout}
\frac{d^2 \ehDistance}{d\TimeVar^2} = \frac{G_N \hM \ehDistance(0)}{\ehDistance(\TimeVar)^3} (1 - \eBetaRad(0)) - \frac{G_N \hM}{\ehDistance(\TimeVar)^2} (1 - \eBetaRad(\TimeVar)) .
\end{equation}
The evolution of orbital distance for a quasite is slow:
\begin{equation}
\ehDistance(\TimeVar) \approx \ehDistance(0) \frac{1 - \eBetaRad(0)}{1 - \eBetaRad(\TimeVar)} \approx \ehDistance(0) \left[1 + \frac{\Delta \eBetaRad(\TimeVar)}{1 - \eBetaRad(0)}\right] .
\end{equation}
In most cases, the main threat would be the dispersion in elements of different $\eSigmaRad$ in closely packed swarms colliding and potentially thermalizing the swarm, because stellar luminosity evolution is so slow. Quasites orbiting a star near the tip of the red giant branch are in a precarious position once it begins its rapid descent as core helium burning begins; the swarm could partly collapse in mere millennia.

True statites can be blown out quite rapidly. Equation~\ref{eqn:eBlowout} is a nonlinear equation with no obvious analytic solution even when $\eBetaRad(0) = 1$, but a very rough sense of the evolution can be had by ignoring the effects of the inverse square law on the acceleration until $\ehDistance(\TimeVar) = (1 + \zeta) \ehDistance(0)$ with $\zeta \sim 1$, beyond which there is no further acceleration. If the luminosity increases linearly, $\eBetaRad(\TimeVar) = 1 + \TimeVar / \hTimeLum$, then the acceleration phase lasts
\begin{align}
\nonumber \TimeVar^{\rm acc} & \sim [2 \pi^2 \zeta]^{-1/3} {\hTimeLum}^{1/3} \ePeriod(0)^{2/3} \\
& \sim 1100\ \yr\,\zeta^{-1/3} \left(\frac{\hTimeLum}{10\ \Gyr}\right)^{1/3} \left(\frac{\ePeriod}{1\ \yr}\right)^{2/3},
\end{align}
at which point the erstwhile statite has reached a ballistic radial velocity of 
\begin{align}
\nonumber \dot{\ehDistance}(\infty) & \sim \dot{\ehDistance}(\TimeVar^{\rm acc}) \sim \left(\frac{9}{2\zeta^2} \frac{G_N \hM}{\hTimeLum}\right)^{1/3} \\
& \sim 0.003\ \AU\,\yr^{-1}\,\zeta^{-2/3} \left(\frac{\hM}{\Msun}\right)^{1/3} \left(\frac{\hTimeLum}{10\ \Gyr}\right)^{-1/3} .
\end{align}
This would lead to a statite swarm in the Solar System being cleared within about $10^5\ \yr$, if the Sun's secular evolution is steady.

Significant radiation pressure support may alter the effects of gravitational perturbations. In terms of Newtonian gravitational attraction, the host star's mass is effectively reduced. A luminous perturbing companion's attractive pull on the element is also reduced, but the amount of the mass reduction depends on the luminosity of the companion and likely varies as the geometry changes during the orbit. But the attraction between the host and the perturber is unchanged. These effects may conspire to give rise to novel dynamical effects.

\subsection{Transverse forces: Yarkovsky effects as perturbations}
\label{sec:Yarkovsky}
In order for the thermal emission of a body to cause net thrust, the temperature of the body must be uneven.\footnote{If the body has \emph{perfectly} uniform emissivity, then the amount of emission in one direction is exactly proportional to its projected area on the transverse plane. However, the projected area is the same when looking at the transverse plane from either side (whether a ray pierces a shape does not depend on whether it comes in from the front or back), so there is no net difference.} The Yarkovsky effect is likely to be the most important source of perturbations. It is the result of the combination of thermal lag and the rotation of a body, which causes the hottest point of a body to be off from the subsolar point. Thus, the radiative thrust from the temperature anisotropy is not perfectly radial; the transverse thrust perturbs the orbit. There are several different manifestations of the Yarkovsky effect, including diurnal and annual variants \citep{Bottke06}. There is an optimal rotation rate that maximizes the diurnal Yarkovsky effect: if it spins too quickly, the body is heated evenly, but if it spins too slowly, the transverse component of the radiative anisotropy is small. A big body is hard to accelerate, but in a small body, thermal conduction redistributes heat efficiently \citep{Bottke06}.

The complicated interplay of geometry and thermal properties makes an exact calculation impossible. However we can set limits on the Yarkovsky effect for sufficiently small objects. Suppose the elements are thin, solid plates with no hollows (e.g., a quasite sail) with a homogeneous thermal conductivity $\eThermConductivity$ and thickness $\eThickness$. Now, the maximum possible temperature anisotropy occurs when the element is in synchronous rotation, so even though asynchronous rotation is necessary for the Yarkovsky effect, we can set a limit on how much thrust is available in this limit. The elements are so thin that all of the thermal emission comes out of the hot, sun-facing side or the cold, sun-shaded side; each side has a uniform temperature, and the difference in temperatures is $\Delta \eTemperature$. From energy balance, the thermal emission from the cold side is equal to the heat flux that can be supported by thermal conduction, $\dot{E}_{\ElementMark} = \eAreaRad (\Delta \eTemperature)/\eThickness$. It can then be shown that, for blackbody surfaces,
\begin{align}
\nonumber \Delta \eTemperature & = \sigmaSB (\eTemperatureHot - \Delta \eTemperature)^4 \eThickness/\eThermConductivity \approx \sigmaSB (\eTemperatureHot)^4 \eThickness/\eThermConductivity \\
& \approx 4.6\ \Kelvin\,\left(\frac{\eTemperatureHot}{300\ \Kelvin}\right)^4 \left(\frac{\eThickness}{1\ \meter}\right) \left(\frac{\eThermConductivity}{100\ \Watt\,\meter^{-1}\,\Kelvin^{-1}}\right)^{-1}
\end{align}
when the temperature difference is much smaller than the temperature of the hot side, $\eTemperatureHot \approx [\hL / (4 \pi \sigmaSB {\ehDistance}^2)]^{1/4} $.

\begin{widetext}
Given these assumptions, the maximum \emph{net} thrust is just $\sim 4 \Delta \eTemperature / \eTemperatureHot$ of the total thrust from all emission. Thus, for a thin blackbody conducting plate,
\begin{align}
\nonumber \eTRadYarkovsky & \ga \frac{\eTemperatureHot}{4 \Delta \eTemperature} \frac{\ePeriod}{2 \pi \eBetaRad} \approx 2^{3/2} \pi^{7/4} \frac{c \eThermConductivity \eRhoRad \sqrt{G_N \hM {\ehDistance}^3 \eSemimajor^3}}{\sigmaSB^{1/4} \hL^{7/4}}  \\
\label{eqn:tYarkovskyPlate}
 & \ga 1.5\ \Myr\,\left(\frac{\hM}{\Msun}\right)^{1/2} \left(\frac{\eSemimajor}{\AU}\right)^3 \left(\frac{\ehDistance}{\eSemimajor}\right)^{3/2} \left(\frac{\hL}{\Lsun}\right)^{-7/4} \left(\frac{\eThermConductivity}{100\ \Watt\,\meter^{-1}\,\Kelvin^{-1}}\right) \left(\frac{\eRhoRad}{\rhoUnits}\right) ,
\end{align}
where $\eRhoRad = \eSigmaRad / \eThickness = \eMass / (\eThickness \eAreaRad)$ is the element's (effective) mass density. The thermal conductivity has been scaled to a value appropriate for iron. This is about a hundred times the thermal conductivity for bare rock \citep{Bottke06}, and so a non-porous metallic body is much less susceptible to the effect than an asteroid of the same size. 
\end{widetext}

Whether the Yarkovsky effect is the dominant source of perturbations over secular gravitational effects depends on how the ETIs would design their elements, and whether they specifically try to compensate for it. Perhaps ETIs have reason to use thermally insulating materials that could sustain much larger thermal inhomogeneities. Heat conduction is hampered if parts of the element are on booms, or if it is largely hollow. On the other hand, heat pipes are able to effect conductivities hundreds of times those of metals without any moving parts, and could be used to even temperature differences.

The main takeaway, however, is that radiative perturbations are a threat close to the star unless the elements are massive, which is also where secular gravitational perturbations and stellar flybys have the least effect.

\section{Implications}
\label{sec:Implications}

\subsection{Where and what are the longest living megaswarms?}
Megastructures may come in a wide range of different types, environments, and purposes. It is possible that in most cases, the builders do not even care about longevity, needing only a quick burst of vast power, although uses like stellar engineering require long-term survival. But if megastructures exist, the population we \editOne{would} observe \editOne{should be} biased towards longer-lived types. So what are the characteristics of these swarms?

\subsubsection{Location of the megaswarm in the system}
If swarm elements were placed in random orbits, there is a clear answer: as far from the host star as possible. Even for a \editOne{fixed} shell \editOne{thickness}, the local number density of elements is, at worst, constant as we move further out. Although the number of elements that is needed grows linearly (minimal occulter) to quadratically (energy collection), the surface area available to spread out grows quadratically. Moreover, the further out one goes, the thicker the shell can be; there is simply more room in the radial direction, and so the shell-averaged number density falls inversely with its radius. Finally, the randomized velocities are slower. This both slows down the rate of collisions, and dampens the cascade by limiting the severity of impacts.

But megaswarms designed for longevity will not be randomized, at least not close to the star, and the relevant timescale is the time to perturb orbits until they cross. There are many trade-offs that make it unclear if there is one optimal location, but placing a swarm close to the host star could be a viable strategy. First, in the base configuration, the fact that there is only one belt at each semimajor axis value means more of the shell is empty as $\sRout$ increases, wasting space. Moreover, the number of belts increases linearly further out; although the linear spacing between belts is constant, the eccentricity needed for belt overlap falls inversely with distance. If element collisions lead to thermalization, then this is the fatal quantity; if thermalization does not occur, however, the eccentricity needed to ensure collisions are destructive grows as we go further out.

The most likely source of gravitational perturbations are secular effects induced by a companion. The worst place for a swarm is then right next to the companion, because of the rapid evolution that results. Exterior to the companion, there are no quadrupole-order effects that raise eccentricity like the Lidov-Kozai effect, but there are still octupole effects if the companion has a nonzero eccentricity. Additionally, stellar flybys are more detrimental further out. Swarms interior to companions have less margin for error and may experience the Lidov-Kozai effect if the range of inclinations is too big. However, the Lidov-Kozai timescale becomes more forgiving closer to the host, and general relativistic precession inhibits the effect entirely. Stellar oblateness may also have a stabilizing effect. This suggests long-lived ordered megaswarms in planetary systems like our own can be placed interior to the radius where $\eTGR \la \epTQuad$. In our own Solar System, there indeed is a stable region interior to Mercury (roughly from $0.04 \endash 0.2\ \AU$), the Vulcanoid zone. 

Radiative perturbations are a real threat to megaswarms orbiting close to the star, but is extremely sensitive to its unknown material properties. The estimates in Section~\ref{sec:Yarkovsky} imply that, within the Solar System, the Yarkovsky effect would dominate of thin elements within the orbit of Venus (Figure~\ref{fig:SolSys}), unless they were specifically designed to redistribute heat (by using heat pipes, for example). Equation~\ref{eqn:tYarkovskyPlate} however sets the \emph{worst}-case limit, however; an asynchronously rotating object redistributes heat and has a smaller radiative thrust. Additionally, massive objects with $\eBetaRad \ll 1$ are resistant to radiative perturbations. Overall, the innermost Solar System appears to be a hostile place for a swarm of small grain-like elements, but might be viable for \editOne{massive swarm bodies}, perhaps including occulter swarms. Placing a swarm that close in may also be more efficient for energy collection \citep{Wright23}. If there are any other sources of perturbation, however, the inner system may be disastrous because of the rapidity of cascades.

For minimal occulter swarms, the outer reaches of the host system may be preferable. Although stellar encounters and very distant companions may randomize orbits, the number density becomes so low that it does not even matter whether the swarm is randomized or not. Even a swarm at $1\ \AU$ is in this regime, and the collisional lifespan increases as $\sRout^{5/2}$: at 10 AU, it is $160\ \Myr$; beyond 40 AU, it is comparable to the lifespan of the Solar System. 

Interestingly, these considerations suggest the habitable zone of a star system may be one of the \emph{least} likely places to find a Dyson swarm, given the large number of warm and cool giant planets that can induce a Kozai-Lidov effect. Megaswarms consisting of habitats also would be so massive that self-driven secular perturbations would destroy them in millennia. If so, searches for waste heat may have better luck looking for uninhabitable megaswarms in the near- or far-infrared.

\subsubsection{Host system properties}
It is more clear what sort of architecture the host system should have to optimize for longevity, namely, \emph{none} -- there should just be the host star and absolutely nothing else in the system. Indeed, perhaps ETIs aggressively re-engineer stellar systems to remove anything that might interfere with their megaswarms. Still, this would not remove the effect of stellar flybys and internal instabilities. Additionally, the materials needed to build the megaswarm may need to be mined by local (solid) planets, or at least minor bodies.

Binary stars are terrible places for megaswarms, unless they are very close together (for a distant swarm) or very far apart (for a close-in swarm). Unbinding stellar companions also takes a long time, even for a Kardashev II \editOne{ETI} -- perhaps longer than a swarm can be maintained. In general, massive stars are more likely to have stellar companions, while red dwarfs are more likely than not to be solo \citep[e.g.,][]{Raghavan10,Winters19}. This suggests that long-lived megaswarms are more likely to be found around low mass stars, particularly cool dwarfs. Note that stars less massive than the Sun put out only a tiny fraction of a galaxy's light; a Kardashev III metasociety would probably at least have to enclose red giants of about Solar mass to capture most of the luminosity \citep{Lacki19-Sunscreen}.

Giant planets are another bane of megaswarms. Again, greater stellar mass implies a greater frequency of giant planets \citep[e.g.,][]{Ghezzi18,Nielsen19}. Here the situation is slightly more complicated: the Lidov-Kozai timescale depends on the mass \emph{ratio} of the planet to the star and the period. The habitable zones of red dwarfs are very close in, with short orbital periods and thus short secular perturbation timescales (equations~\ref{eqn:tQuadrupole}--\ref{eqn:tOctupole}). A more useful discriminant may be stellar metallicity, which correlates with the presence of giant planets: low metallicity stars are more likely to lack these destructive neighbors \citep[e.g.,][]{Johnson10}. A minimum metallicity threshold of is suspected to be necessary for terrestrial planet formation\editOne{:} $[\mathrm{Fe/H}] \sim -2$  \editOne{is assumed in \citet{Zackrisson16}, $\sim -1.5$ is proposed in \citet{Johnson12} (dependent on distance), and up to $\sim -0.6$ in \citet{Andama24}. G}oing below that can be helpful in prolonging the swarm's existence, although probably not for building it in the first place. Minor body impacts may also still be an issue in these systems, shattering the first elements and triggering the cascade.

The close-in habitable zones of red dwarfs pose additional problems. At a fixed \editOne{instellation} $\ehInsolation$, orbital velocities are higher, as $\xVCirc \propto (\ehInsolation)^{1/4} \hM^{1/2} / \hL^{1/4}$. Red dwarfs, for example, have orbital velocities about twice as high as those around the Sun at a given \editOne{instellation} (Table~\ref{table:CollisionTimes}). The problem gets even worse for cool white dwarfs. For these low luminosity stars, cold megaswarms shining in the far-infrared may be the norm.

High mass stars also have a disadvantage: they just don't live very long, at least on cosmological timescales. Their rapid evolution can especially be an issue for statites levitated by radiation pressure; as their main sequence luminosity grows, the elements are ejected from the system.

The collisional destruction of megaswarms also may have implications for where in a galaxy we might find them. First, stellar encounters should be minimized, implying low density environments. These are found on the outskirts of large galaxies and in their halos, as well as in dwarf galaxies. Second, low metallicity stars are more likely to be found in these same environments. This suggests that megaswarms are more likely to be found in regions that are sometimes considered disfavorable for habitability (compare with \citealt{Lineweaver04} and subsequent works).

\subsection{Self-sabotage and the end of planets?}
\label{sec:PlanetDestroyers}

The more aggressive option is to remove the intruding perturbers, either by  ejecting them from the system or destroying them outright. For objects with a smaller surface escape velocity than orbital velocity, which mainly includes terrestrial planets, it is energetically favorable to dismantle them entirely; otherwise, it may be easier just to unbind them \editOne{from the host}, as for stellar companions and most giant planets. The amounts of energy needed are astronomical, but \editOne{for planetary perturbers,} well within reach for a true Kardashev type II \editOne{ETI}, needing a few days of Solar luminosity to unbind an Earth and decades to eject a Jupiter. Nonetheless, either option is very difficult for stellar companions; it would take many millennia to eject a star, and in the meantime, that same star is destabilizing the megaswarm. 

Circularizing orbits may be another option if the builders are content with a swarm that remains within the range of inclinations that are Lidov-Kozai stable. In the Solar System, where most of the major planets already have low eccentricities, this is easier than removing the planets. Nonetheless, the planets may need to be occasionally recircularized against their mutual influences and the effects of stellar flybys. Other stellar systems have eccentric companions, however; eccentricities of $\sim 0.5$ are common among binary stars with periods greater than 100 days \citep{Raghavan10}, in which case the needed $\Delta \xV$ is only perhaps a factor $\sim 2$ smaller than ejection.

This raises a disturbing prospect -- a Kardashev Type III metasociety may annihilate planetary systems as it expands to stabilize their megaswarms, leaving everything within its ``bubble'' barren of planets. If those megaswarms then are destroyed anyway, no planets will remain on which to restart life. Furthermore, in most early-type galaxies, there will be no star formation to ever replace them. Thus, the megaswarms portend doom, and leave behind a permanently sterilized galaxy in their wake. Also note that the perturbation times are generally comparable to or shorter than the canonical $1 \endash 100\ \Myr$ expected to expand across a galaxy. It could be that at no point there is a galaxy entirely shrouded by Dyson swarms, only a thin front.

\subsection{Another option: maintained swarms}
Of course, all of these considerations assume that the megaswarm is abandoned to its own devices. For all we know, the builders are necessarily long-lived and can maintain an active watch over the elements and actively prevent collisions, or at least counter perturbations. Conceivably, they could also launch tender robots to do the job for them, or the swarm elements have automated guidance. Admittedly, their systems would have to be kept up for millions of years, vastly outlasting anything we have built, but this might be more plausible if we imagine that they are self-replicating. In this view, whenever an element is destroyed, the fragments are consumed and forged into a new element; control systems are constantly regenerated as new generations of tenders are born. Even then, self-replication, repair, and waste collection are probably not perfectly efficient: very small grains are blown out of the system by radiation pressure, adding a leak to the closed system. These may be supplemented by mining of solid bodies, although one might wonder if self-replicating probes that feast on these might evolve to ignore the megaswarm and let it be destroyed.

Another issue is that orbital traffic control generally requires propellant. \citet{Wright20} noted the issue for stabilizing a solid shell around a star against the effect of a perturber. According to the Tsiolkovsky rocket equation, the amount of propellant required grows exponentially for a $\Delta \xV$ beyond the exhaust velocity, and since the amount of power available is limited by the star's luminosity, the swarm's mass is eventually consumed. \citet{Wright20} estimated this lifespan to be of order centuries for a Jupiter-mass solid shell being perturbed by a Jupiter-mass planet, with less massive shells surviving longer. For a megaswarm, the conditions are not so stringent. We merely need to suppress the perturbations, nudging the elements back onto circular orbits. If the swarm is being actively maintained, these nudges might be more like a continuous thrust. There is a continuous source of propellant in the host star's light, which can be reflected to counter the perturbations. From momentum conservation, the amount of delta-$v$ that an element can extract from starlight is limited to 
\begin{align}
\nonumber \xDeltaV & \approx \frac{\hL \eTPerturb}{4 \pi c \sRout^2 \eSurfDensity} \\
& \approx 0.14\ \kms \left(\frac{\hL}{\Lsun} \cdot \frac{\eTPerturb}{\kyr}\right) \left(\frac{\sRout}{\AU}\right)^{-2} \left(\frac{\eSigmaRad}{100\ \SigmaUnits}\right)^{-1}
\end{align}
where $\eTPerturb$ is the timescale of the perturbations. For comparison, the correction needed to circularize an eccentric orbit is $\sim \eEccentricity \xVCirc \approx 0.3\ \kms (\eEccentricity/0.01) (\xVCirc/30\ \kms)$. This suggests that starlight alone may be sufficient to counter perturbations in a system like our own, given that these perturbations take tens of thousands of years to develop. If a Lidov-Kozai instability is allowed to fully develop, the eccentricity swells to $\sim 1$ and sunlight is insufficient, but note that the idea here is that the element is constantly correcting the perturbations, squelching any changes in eccentricity much like general relativistic precession does.

In fact, propellant may not be needed at all. For every collision, there are bound to be many near-misses. The elements may take advantage of these near-misses to grapple onto each other with tethers and fling each other apart, exchanging momentum to provide propulsion. This is especially easy in a dense swarm, where the first encounters are likely to be very slow.

It all comes down to the original question from the introduction: can technological societies \editOne{(or their equivalents)} survive for millions of years? Or can they at least create autonomous agents that maintain their technosignatures for that long? If they can, then they can presumably maintain any technosignature; the advantage of megaswarms is simply that they are detectable from very far away. If they cannot, then megaswarms around stars are not eternal monuments; they too shall perish in time, leaving no trace of their makers.

\subsection{Galactic-scale megaswarms as the survivors}
Stellar megaswarms, without maintenance, are expected to be destroyed in most cases within a few million years -- that is, within about the length of time expected for an ETI to spread across a galaxy. The archetypal Kardashev III ETI is a galaxy wherein all the stars have been shrouded in Dyson spheres. But, supposing that such structures are destroyed so quickly, is the very idea of a Kardashev III ETI implausible, with a wave of settlement immediately trailed by a wave of destruction, leaving only a thin skein of infrared glow expanding through the galaxy?

Not quite. 

The problem with stellar megaswarms is that the most fundamental dynamical timescale -- the orbital period -- is so short. Hence the need for careful structuring of orbital belts to ensure that they do not cross, and delaying the cascade until perturbations kick in. But the dynamical timescale of a galaxy \emph{as a whole} is vastly longer, more like a hundred million years. 

A megaswarm that covers a galaxy thus could plausibly survive passively for many millions of years. Although the material requirements may seem too vast, large star-forming galaxies in fact achieve optical thickness of $\sim 0.5 \endash 1$ to visible light, a result of interstellar dust \citep{Calzetti01}. Dust in the interstellar medium gets around several issues that plague stellar megaswarms. First, the dust-to-gas ratio is $\sim 1\%$, much higher than the planet-to-star (to say nothing of the solid-to-star) ratio of the Solar System. Second, dust grains are very small and can be spread out to cover a wider area. Unlike Dyson swarms, however, the dust grains are \emph{weighed down by the interstellar medium}, which prevents the radiation pressure from instantly blowing away the dust. Of course, a galaxy must have enough of an interstellar medium in the first place for this advantage to occur; a majority of early-type galaxies lack detectable cool gas, though it is not a universal condition \citep{Young14}. The most extreme example of a galactic megaswarm is a ``blackbox'', in which dipole antennas are used to reprocess the starlight into microwaves, turning the galaxy into a blackbody. These are near the mass limits in the interstellar medium (ISM) of a star-forming galaxy \citep{Lacki16-K3}.

The presence of the interstellar medium is also likely to alter the evolution of galactic dust megaswarms. For one, gas drag will cause the artificial grains to be more or less locked into the ISM flows, suppressing local velocity gradients, possibly preventing collisions from destroying grains. On the other hand, some ISM environments may destroy grains -- large volumes of hot ISM could spallate away the artificial grains \citep[as with natural grains, see][]{Draine11}.

Although there remain many unexplored issues about the feasibility and evolution of such galactic structures \citep{Lacki16-K3}, it remains a possibility that they are the only persistent technosignature of a Kardashev III ETI, perhaps the only way to simultaneously capture the light of all stars in a galaxy.

\section{Conclusions}
\label{sec:Conclusions}
At first glance, megaswarms may seem to be ideal passive technosignatures, eonic structures that sit invulnerable as the universe around them ages. This paper deals with the threat posed by collisional cascades for megaswarms surrounding stars. These swarms, practically by definition, need to have a large number of elements, whether their purpose is communication or exploitation. Moreover, the swarm orbital belts need to have a wide range of inclinations. This ensures that the luminosity is being collected or modulated in all directions. But this in turn implies a wide range of velocities, comparable to the circular orbital velocity. Another problem is that the number of belts that can ``fit'' into a swarm without crossing is limited.

The collisional time for a megaswarm is very short. For a Dyson swarm, it is as small as a single orbital period divided by the covering factor. Even for a ``minimal occulter'' swarm, the collision time can be less than a million years. This is true even if the swarm is structured by imposing circular orbits on all elements and placing elements of similar inclinations on similar orbits. To avoid high-velocity collisions, most of the swarm's volume must be empty, compensating for the reduced velocity differences in the collision time. Nonetheless, this kind of ordering is advantageous: the cascade time can be \emph{much} shorter still than the collisional time. My cascade simulations indicate a $\xTCasc \propto \Mean{\eeVRel}^{-8/3}$ dependence. Thus, I argue that any megaswarm built to last has the ``base configuration'' of this paper, involving a series of circular orbital belts with a gradual change in inclination with semimajor axis. 

Unless the megaswarm is actively maintained, gravitational and radiative perturbations cause precession and, in some cases, eccentricity growth. These cause the belts to puff out and overlap, allowing for collisions of elements both within and between belts. Even if the collisions between elements are too slow to destroy the elements, the momentum transfer could cause the swarm to thermalize, a disastrous effect given the wide range of inclinations in the swarm. There are many sources of such perturbations, although some are relatively ineffective. Among the most powerful are quadrupole and octupole secular perturbations from companion stars and planets. The most famous manifestation is the Lidov-Kozai effect, which turns circular orbits at high inclination into eccentric orbits at low inclination. I have shown that the timescales for many such systems is in the tens of millions of years or less, often much less. Jupiter, for example, would destroy a megaswarm at Earth's place in a few hundred thousand years. To prevent this from happening, the builders must actively maintain the megaswarm -- or remove the perturbers by ejecting them from the system or destroying them outright. This raises the disturbing \editOne{possibility} that ETIs might sacrifice all planetary systems in a galaxy, leaving no foothold for life to recover if they do collapse. Stellar oblateness (particularly around early-type stars) and general relativity are sources of precession, randomizing already elliptical orbits, but they suppress eccentricity growth. Massive swarms have their own internal Lidov-Kozai-like perturbations. Encounters with passing stars raise eccentricity, though it is slow even on geological timescales except in the outermost regions of planetary systems.

Radiation pressure from the host sun is another source of perturbations, and it is fundamentally dispersive because it will affect elements in the same orbit differently according to their compositions and rotations. Although it can be used to partially levitate swarm elements and reduce their collision speeds, stellar variability, including starspots, will introduce randomness into the swarm element orbits. Transverse radiative forces resulting from the Yarkovsky effect can be expected to be major source of perturbations, although to a very uncertain degree. Small elements might suppress the Yarkovsky effect if they have a high thermal conductivity, evening out temperature differences; the perturbation time might then be of order a few hundred thousand years.

\editOne{The final result of a collisional cascade is to grind a swarm down into microscopic particles. These artificial grains are blown out into the general interstellar medium if the host star is bright enough. Otherwise, the swarm is ultimately reduced into ionized gas.} 

There remain significant issues for future work. Is the configuration of circular orbits with gradually increasing inclination the best for resisting cascades? Simulations of cascades that take spatial configuration into account might provide insight into how long the swarm really lasts. \editOne{A more realistic treatment of an evolving velocity distribution for the swarm fragments, accounting for the effects of dissipation, is warranted.} The perturbations can also be studied in more detail. How does radiation pressure support affect the Lidov-Kozai and similar effects, and especially, how quickly does it introduce perturbations into realistic satellites? What if there are multiple perturbing companions?

The destruction wrought by collisional cascades are a general property of orbital swarms, no matter the scale.  The process has likely already begun in Earth orbit even with our sub-Kardashev I efforts. Just as land on Earth is finite, so too is the phase space of Earth orbit, the Solar System, the Galaxy, and the Universe, orders greater it may be. If ETIs keep expanding, eventually they reach these limits. They then must either coordinate and regulate these commons, work without cease to repair what they have built, or let it all be ground to dust.

\acknowledgments
{I thank the Breakthrough Listen program for their support. Funding for \emph{Breakthrough Listen} research is sponsored by the Breakthrough Prize Foundation.\footnote{\url{https://breakthroughprize.org/}} I am grateful to Jason Wright for reading the manuscript and providing interesting discussions\editOne{, especially about the role of radiation pressure}. \editOne{I also thank the referee for comments regarding dissipation in the cascade.} In addition, I acknowledge the use of NASA's Astrophysics Data System and arXiv for this research.}

\bibliographystyle{aasjournal}
\bibliography{GroundToDust_arXiv_v2}

\end{document}